\documentclass{aa}

\usepackage{graphicx}
\usepackage{caption}
\usepackage{subcaption}
\usepackage{epstopdf}
\usepackage{amsmath}
\usepackage[english]{babel}
\usepackage{amssymb}
\usepackage{wrapfig}
\usepackage{pdflscape}
\usepackage{lscape}
\usepackage{longtable}
\usepackage{appendix}
\usepackage{textcomp}
\usepackage{multirow}
\usepackage{array}
\usepackage{ragged2e}
\usepackage{adjustbox}
\usepackage{txfonts}

\begin{document} 

   \title{Near-infrared morphologies of the host galaxies of narrow-line Seyfert 1 galaxies}


   \author{E. J\"{a}rvel\"{a}\inst{1}\fnmsep\inst{2}\thanks{\email{emilia.jarvela@aalto.fi}}
          \and 
          A. L\"{a}hteenm\"{a}ki\inst{1}\fnmsep\inst{2}\fnmsep\inst{3}
          \and
          M. Berton\inst{4}\fnmsep\inst{1}\fnmsep\inst{5}\fnmsep\inst{6}
          }

   \institute{Aalto University Mets\"{a}hovi Radio Observatory, Mets\"{a}hovintie 114, FI-02540 Kylm\"{a}l\"{a}, Finland
         \and
             Aalto University Department of Electronics and Nanoengineering, P.O. Box 15500, FI-00076 AALTO, Finland
        \and
             Tartu Observatory, University of Tartu, 61602 T\~{o}ravere, Tartumaa, Estonia
        \and
            Finnish Centre for Astronomy with ESO (FINCA), University of Turku, Quantum, Vesilinnantie 5, 20014 University of Turku, Finland
        \and
             Dipartimento di Fisica e Astronomia ”G. Galilei”, Universit{\`a} di Padova, Vicolo dell’Osservatorio 3, 35122 Padova, Italy
        \and 
             INAF - Osservatorio Astronomico di Brera, via E. Bianchi 46, 23807 Merate (LC), Italy
             }

   \date{Received ; accepted }


  \abstract
   { 
   We present $J$-band near-infrared (NIR) imaging of the host galaxies of nine narrow-line Seyfert 1 galaxies (NLS1). Based on high-frequency radio observations at 37~GHz, seven of them could host powerful, most likely relativistic jets. 
   Host galaxy morphology studies of NLS1 galaxies are scarce, but exceedingly important for understanding the seemingly heterogeneous nature of the NLS1 population as well as their evolution and place in the active galactic nuclei (AGN) scheme. Increasing the sample size is essential for achieving statistically significant results.
   We determine the morphological types of the host galaxies by performing photometric decomposition of NIR images using a 2D image decomposition algorithm GALFIT.
   We were able to sufficiently model five of the nine host galaxies. Based on the fitting parameters, mainly the S\'{e}rsic index, all of them are disk-like galaxies. Sources with clearly distinguishable bulge components all have pseudo-bulges, and four out of five sources show a component resembling a bar. A surprisingly large fraction, three out of five, show signs of interaction or disturbed morphology.
   Our results suggest that spiral galaxies with pseudo-bulges are able to launch and maintain powerful jets. They also imply that interaction ---mainly minor mergers--- may have a role in initially triggering higher levels of nuclear activity in NLS1 galaxies. Furthermore, our results support the heterogeneous nature of the NLS1 class and indicate that this diversity is caused by different evolutionary stages, possibly due to mergers.
   }
   

   \maketitle


\section{Introduction}
\label{sec:intro}

Narrow-line Seyfert 1 galaxies \citep[NLS1, ][]{1985osterbrock1} are a subclass of active galactic nuclei (AGN), characterised by the narrow permitted and forbidden emission lines \citep[$FWHM$(H$\beta$)$ < 2000$ km s$^{-1}$,][]{1989goodrich1}. In addition, [O~III] $\lambda$5007 emission is relatively weak \citep[F($\lbrack$O~III$\rbrack$)/F(H$\beta) <$ 3,][]{1985osterbrock1} and Fe~II emission relatively strong in the majority of NLS1 sources \citep{1985osterbrock1}. Several NLS1 galaxies have been detected at gamma-rays \citep[{e.g.},][]{2009abdo2,2018paliya1}, confirming the presence of powerful relativistic jets in them.

In contradiction to the usual view that only supermassive black holes ($M_{\text{BH}} > 10^{8}$ $M_{\sun}$) are able to launch powerful relativistic jets, \citep{2000laor1} the black holes in NLS1 galaxies have low or intermediate masses \citep[$M_{\text{BH}} < 10^{8}$ $M_{\sun}$, ][]{2000peterson1} and accrete at extraordinarily high rates, from 0.1 Eddington ratio to even super-Eddington accretion \citep{1992boroson1}. However, it has been suggested that the unusually low black hole masses, when estimated using the emission line widths, are due to the orientation effects caused by a flattened broad-line region (BLR) seen face-on \citep{2008decarli1,2017rakshit1}. This would especially affect the gamma-ray detected sources, known to have small inclinations. Alternative methods to estimate the black hole masses, for example, the $M_{\text{BH}}$ -- $L_{\text{bulge}}$ relation \citep{2007ryan1,2018dammando1}, spectropolarimetry \citep{2016baldi1}, and accretion disk modelling \citep{2013calderone1}, result in higher masses. On the other hand, mass estimates based on line dispersion, which is less affected by the orientation \citep{2015foschini1,2018berton1}, and reverberation mapping \citep{2013rafter1,2016wang1} favour low black hole masses.

Only $\sim$10$\%$ of NLS1 sources are radio-loud (radio-loudness{\footnote{Radio-loudness, $R$, is defined as the ratio between radio flux density, $S_{R}$, and optical flux density, $S_{O}$, originally at 5~GHz and in $B$-band, respectively \citep{1989kellermann1}.}} $R$ $>$ 10) \citep{2006komossa1}, $\sim$6$\%$ radio-quiet ($R <$ 10), and the rest have not been detected in radio at all and are thus radio-silent. Radio-loudness particularly at lower frequencies does not necessarily indicate the presence of a jet, and on the other hand, some radio-silent NLS1 sources have recently been detected at 37~GHz \citep{2018lahteenmaki1}, further highlighting the ambiguity of the radio-loudness parameter. Diverse radio properties and other disparities, for example, in the black hole masses and the large-scale environments, suggest that the NLS1 population is not homogeneous. It remains unclear which intrinsic properties account for the various characteristics observed in them, and how --- if at all --- these seemingly different NLS1 subpopulations are connected.

The connection between an AGN and its host galaxy has already been established \citep[e.g.,][]{2008storchi-bergmann1,2010vandeven1,2012povic1,2012fabian1,2015king1}. The host has a direct impact on the activity of the AGN by regulating the gas supply of the central black hole. In turn, various AGN feedback mechanisms, for example, radiation pressure, jets, and winds and outflows, affect the host galaxy and steer its evolution. Feedback can be either negative or positive, for example, quenching or enhancing star formation.

The nearby environment of a galaxy could affect its properties and subsequently the nuclear activity. A dense enough environment increases the probability of galaxy -- galaxy interaction, and a galaxy might undergo multiple minor and major mergers, and close encounters, that distort and change its morphology. Interaction and mergers might replenish or strip the gas reservoirs of the galaxy and cause gas infall, which in turn triggers circumnuclear star formation and feeds the black hole. A clear connection between a merger and the level of AGN activity is yet to be established. Some studies argue that both minor \citep{1999taniguchi1,2008barth1} and major \citep{2008urrutia1,2011ellison1} mergers can trigger nuclear activity, whereas others do not find mergers and nuclear activity connected at all \citep{2000corbin1,2011cisternas1,2012kocevski1}. It has even been suggested that all radio-loud AGN are mergers \citep{2015chiaberge1}, predominantly major, or that the most luminous AGN are mostly triggered by major mergers \citep{2012treister1,2014villforth1}.

Host galaxy studies of NLS1 galaxies have so far mostly concentrated on the radio-quiet and radio-silent sources. They are preferably, but not exclusively, found in disk-like galaxies. Large-scale stellar bars are frequent (65\%--80\%) in NLS1 galaxies \citep{2003crenshaw1,2007ohta1}, as are nuclear dust spirals (83\%) of which most are grand-design (80\%) \citep{2006deo1}. NLS1 galaxies also show enhanced star formation in their nuclear regions \citep{2006deo1,2010sani1}. \citet{2000mathur1} suggested that NLS1 galaxies could be sources rejuvenated by a recent merger, but since only 8\%--16\% of NLS1 galaxies show signs of interaction or merging \citep{2007ohta1} \citep[20\%--30\% for Seyfert galaxies in general, ][]{2001schmitt1} this scenario seems insufficient to explain the nuclear activity in all NLS1 galaxies. The paucity of interaction and mergers \citep[e.g., ][]{2007ryan1, 2007ohta1} indicates that NLS1 galaxies mainly grow and evolve through secular processes. The predominance of pseudo-bulges in NLS1 host galaxies further supports this scenario \citep{2011orbandexivry1}. A singular ensemble of properties such as bars, grand-design dust spirals, and stellar nuclear rings, might enable efficient fueling of the central black hole and in part explain extraordinarily high Eddington ratios and enhanced star formation in NLS1 galaxies.

So far the host galaxies of only four radio-loud NLS1 sources have been investigated, all of which have also been detected at gamma-rays. The morphology of 1H0323+342 is unclear. \citet{2007zhou1} claim that it is a one-armed spiral galaxy, whereas others find results indicating it is an elliptical galaxy and disturbed by merging \citep{2008anton1,2014leontavares1}. \citet{2017olguiniglesias1} found FBQS~J1644+2619 to reside in a barred lenticular galaxy showing signs of a recent minor merger, but \citet{2017dammando1} argue that the host is an elliptical galaxy. PKS~2004-447 seems to reside in a barred disk-like galaxy \citep{2016kotilainen1}, and PKS 1502+036 in an elliptical galaxy \citep{2018dammando1}.

Comparing the radio-loud NLS1 host galaxy, multifrequency, and environment data with the corresponding data for radio-quiet NLS1 galaxies and, eventually, other AGN classes, will help us to address several unsolved issues. We will have an additional perspective on the heterogeneity of the NLS1 population and can estimate how significant the host galaxy properties are for the nuclear activity. This  helps us to unveil the mechanism responsible for the triggering and maintaining a relativistic jet. We can also investigate the intraclass evolution of NLS1 sources, especially whether the seemingly different NLS1 galaxies are connected through evolution, or have separate evolutionary lines. A better understanding of their diversity helps us to clarify their place in the AGN unification schemes and aids in the search for the parent population. Additionally, we get insight into the future evolution of NLS1 galaxies and their connection to other AGN classes, for example, flat-spectrum radio quasars, whose scaled-down young versions NLS1 galaxies are assumed to be. NLS1 sources offer us a view to the first stages of the evolution of the most powerful AGN.

In this paper we investigate the near-infrared morphologies of nine NLS1 galaxies. The paper is organised as follows: we explain the sample selection, observations and data reduction in Section~\ref{sec:obsandred}, and the data analysis methods and the results for  individual sources fitted reliably in Section~\ref{sec:dataanalysis}. Compromised fits are presented in appendix~\ref{sec:compromised-fits}. We discuss the results and their implications in Section~\ref{sec:disc}, and conclude in Section~\ref{sec:concl}. Throughout the paper we assume a cosmology with H$_{0}$ = 73 km s$^{-1}$ Mpc$^{-1}$, $\Omega_{\text{matter}}$ = 0.27 and $\Omega_{\text{vacuum}}$ = 0.73 \citep{2007spergel1}.

\section{Observations and data reduction}
\label{sec:obsandred}

Our original sample consists of 13 NLS1 galaxies included in the Mets\"{a}hovi NLS1 observing programme \citep{2017lahteenmaki1,2018lahteenmaki1}, with a redshift limit of $z<0.3$ to ensure good enough quality of NIR data under favourable conditions. The sample is diverse: it includes radio-silent, radio-quiet, and radio-loud sources, both flat- and steep-spectrum sources, and sources detected and not detected at 37~GHz. Our purpose was to as widely as possible probe the host galaxy characteristics of a varied ensemble of NLS1 galaxies, and how they possibly affect the formation of jets in them.
Due to challenging weather conditions we were able to observe nine sources. Basic properties of the sample and the observations are listed in Table~\ref{tab:sample}. Exposure times of the sources vary because we were not able to complete the whole observing programme due to the weather. Details of individual sources and their modelling are provided in Section~\ref{sec:dataanalysis}. 

\begin{table*}
\caption[]{Basic properties of the sample and $J$-band observations.}
\centering
\begin{tabular}{l l l l l l l l}
\hline\hline
Source                   & z      & Scale   & RA          & Dec         & Exp. time & Obs. dates    & Seeing\\
                         &        & (kpc/") & (J2000.0)   & (J2000.0)   & (s)       &                & (")  \\ \hline
SDSS J091313.73+365817.2 & 0.1073 & 1.886   & 09:13:13.72 & 36:58:17.23 & 3600      & March 3-4, 2017 & 0.7 \\          
SDSS J111934.01+533518.7 & 0.1060 & 1.886   & 11:19:34.04 & 53:35:18.11 & 1800      & March 3, 2017 & 0.7 \\          
SDSS J122749.14+321458.9 & 0.1368 & 2.327   & 12:27:49.15 & 32:14:59.04 & 1800      & March 4, 2017 & 1.2 \\  
SDSS J123220.11+495721.8 & 0.2619 & 3.902   & 12:32:20.11 & 49:57:21.75 & 3600      & March 3, 2017 & 1.0 \\
SDSS J125635.89+500852.4 & 0.2453 & 3.717   & 12:56:35.87 & 50:08:52.54 & 3600      & March 3, 2017 & 0.9-1.4 \\    
SDSS J133345.47+414127.7 & 0.2252 & 3.485   & 13:33:45.47 & 41:41:27.66 & 3600      & March 3, 2017 & 1.1 \\   
SDSS J151020.06+554722.0 & 0.1497 & 2.511   & 15:10:20.06 & 55:47:22.04 & 3600      & March 3, 2017 & 1.3-1.7 \\
SDSS J152205.41+393441.3 & 0.0766 & 1.394   & 15:22:05.41 & 39:34:40.71 & 3600      & March 3, 2017 & 1.2 \\        
SDSS J161259.83+421940.3 & 0.2331 & 3.577   & 16:12:59.84 & 42:19:40.32 & 3600      & March 3, 2017 & 0.8 \\  \hline
\end{tabular}
\label{tab:sample}
\end{table*}

\begin{table*}
\caption[]{Basic properties of the PSF stars.}
\centering
\begin{tabular}{l l l l l l}
\hline\hline
PSF star                    &  RA         & Dec          & $J$-band & Seeing \\
                            & (J2000.0)   & (J2000.0)    & (mag)    & (") \\ \hline
TYC 2499-339-1              & 09:13:05.86 & +36:54:51.06 & 10.99    & 0.7 \\
TYC 2528-1329-1             & 12:29:04.37 & +32:16:45.42 & 11.48    & 1.2 \\
TYC 3458-1053-1             & 12:31:09.17 & +50:13:40.23 & 11.11    & 1.0 \\
TYC 3461-2198-1             & 12:57:15.54 & +49:54:14.06 & 10.51    & 0.9 \\
TYC 3029-1092-1             & 13:34:00.47 & +41:26:40.47 & 10.38    & 1.1 \\
TYC 3871-106-1              & 15:10:12.75 & +55:44:53.50 & 11.15    & 1.3 \\
TYC 3052-658-1              & 15:22:10.69 & +39:42:47.91 & 10.63    & 1.2 \\
TYC 3064-693-1              & 16:13:00.64 & +42:26:53.88 & 11.93    & 0.8 \\ \hline
\end{tabular}
\label{tab:psfstars}
\end{table*}

$J$-band observations were performed with the Nordic Optical Telescope (NOT) during four nights, March 3-7, 2017, using the NOTCam with high resolution imaging. The NOTCam CCD is 1024 x 1024 pixels, and with the high resolution imaging the field of view (FOV) is 80"x80", giving a scale of 0.078"/px. The weather and seeing during the observations were variable; we report them separately for each source. 
The total exposure times for each source are shown in Table~\ref{tab:sample}. We used 9-point dithering with frame mode, with the dithering step between 10"--15", depending on the size of the source, and skew of 1". Each $J$-band ($\lambda_{central}$ = 1.247 $\mu$m) observation took 900s (9$\times$ 10$\times$10s). In addition to the targets we observed a bright ($J$-band magnitude $\sim$11) nearby star for each source to properly model the point spread function (PSF). This is needed to properly perform the deconvolution and to get rid of the AGN contamination to allow the modelling of the host galaxy. PSF stars were carefully observed so that they have very high quality, S/N$>$1000, but not saturated. To perform the photometric calibration we observed the standard field AS16 \citep{1998hunt1}.

We reduced the data using the NOTCam data reduction package{\footnote{http://www.not.iac.es/instruments/notcam/}} for IRAF\footnote{IRAF is distributed by the National Optical Astronomy Observatories, which are operated by the Association of Universities for Research in Astronomy, Inc., under cooperative agreement with the National Science Foundation.}. The bad pixel mask provided was used in flat-field and actual images. Morning and evening flats were taken every day and combined to provide a masterflat. Dithered images were aligned using point-like sources in the FOV, and combined to a single image. Since we have multiple exposures of each source, these images were aligned and co-added to form the final image used in the data analysis. The data reduction and image combination were performed similarly for the PSF stars and standard field images.

\section{Data analysis}
\label{sec:dataanalysis}

\subsection{Image analysis}

We used the 2D fitting algorithm GALFIT version 3 \citep{2010peng1} to perform a photometric decomposition of the images. The fitting procedure was similar for all sources. A crucial part of the fitting is to correctly model the PSF and thus be able to remove AGN contamination. For each source we used a separate PSF star with very high S/N, observed right after the source to have as similar conditions, for example, seeing and airmass, as possible. After manually centering the PSF image and subtracting the mean sky value --  which we estimated with IRAF -- GALFIT can extract the PSF directly from the image. This is the most accurate way to model the PSF, but it does not give an analytical form for the subtracted PSF. To ensure as good a spatial resolution as possible we observed the galaxies using the high resolution camera. However, the FOV of NOTCam is so small that generally there are no stars in the same FOV as our targets. Therefore we could not test the goodness of the PSF or estimate how using a separate PSF star affects the results. In most cases the conditions stayed similar during the PSF star and the NLS1 galaxy observations, but in some cases they slightly changed. These are mentioned for individual sources in Section~\ref{sec:individual}.

Before fitting the galaxy the sky background was estimated using IRAF. For each source the sky value was measured in several separate, isolated regions of $\sim$150$\times$150~px. The mean value was used in the fitting. The errors were estimated by fitting with $\pm$1$\sigma$ values. For magnitudes an additional zeropoint estimation error was added. We first fitted the source only with the PSF, and checked the residual image to see whether there was emission left and need for additional components. Then, if needed, we added more components, convolved with the PSF, one at a time, and after each addition checked the residual image for leftover flux and the subcomponent images to ascertain that the model components were physically reasonable. At each step we also varied the initial fitting parameters to make sure that the parameter values the fit converges to are correct and stable. We kept all the parameters free to have as unbiased a fit as possible, except in a few cases we had to freeze the central positions of some components. We used as large a fitting region as reasonably possible (but a smaller region is shown in the images for easier inspection) and tested whether the size of the fitting region affects the results. If there were any other sources inside the fitting region, we also fitted them to achieve better results. We varied the number of components and their parameters until the best, yet reasonable, fit was achieved. The goodness of fit was determined based on the $\chi^2_{\nu}$ of the fit as well as visual inspection of the model components. We mainly used the S\'{e}rsic profile to model the various components of our sources:

\begin{equation}
    I(r) = I_e \texttt{exp} \Bigg[ -\kappa_n \Bigg( \bigg( \frac{r}{r_e} \bigg)^{1/n} -1 \Bigg) \Bigg]
\end{equation}

where $I(r)$ is the surface brightness at radius $r$, $\kappa_n$ is a parameter connected to the S\'{e}rsic index, $n$, so that $I_e$ is the surface brightness at the half-light radius, $r_e$ \citep{2005graham1}. S\'{e}rsic profile can be used to model a variety of different light distributions in galaxies, for example, classical and pseudo-bulges, and elliptical and spiral morphologies, merely by changing the S\'{e}rsic index $n$. Setting $n$=4 gives the traditional de Vaucouleurs profile, widely used to model classical bulges, $n$=1 represents the exponential profile, useful for modelling disks, and $n$=0.5 gives a Gaussian profile. In general, smaller values of $n$ --- a core flattening faster at $r < r_e$ and the intensity dropping faster at $r > r_e$ --- are associated with galaxies with disk-like morphology and pseudo-bulges, and larger values of $n$ with elliptical galaxies and classical bulges. 

For the special case when $n$=1, the functional form of the S\'{e}rsic profile is for historical reasons given as

 \begin{equation}
     I(r) = I_0 \texttt{exp} \bigg( \frac{r}{r_s} \bigg)
 \end{equation}

where $I(r)$ is the surface brightness at radius $r$, $I_0$ is the central surface brightness, and $r_s$ is the scale length of the exponential disk. Explicitly when $n$=1 the scale length is related to the effective radius by $r_e$ = 1.678 $r_s$.

After achieving a sufficient fit we extracted the radial surface brightness profile from the observed image, the model image, and the separate component images using IRAF task $ELLIPSE$, which fits concentric elliptical isophotes to a 2D image. $ELLIPSE$ is given initial values for the central $x$ and $y$ coordinates, ellipticity ($\epsilon$), and the position angle (PA). The ellipticity and the position angle of the subsequent ellipses can vary. Output from this task includes the surface brightness ($\mu$), semimajor axis length, PA, and $\epsilon$ of each isophote. We compare the radial surface brightness profiles of the galaxy and the best fit model using $ELLIPSE$ with similar initial values, because the PSF does not have an analytical form to be plotted. It should be noted that the individual components in the surface brightness profiles are not necessarily concentric with the observed and model images. The errors are estimated taking into account the most significant error sources: the error in sky value estimation, and the error in the magnitude estimations of the stars used to calculate the magnitude zeropoint.

We used an additional diagnostic whenever the fit indicated the presence of a bar. When plotting the ellipticity and the position angle of the isophotes, given by $ELLIPSE$, as functions of radius, a bar can be identified as a region with increasing ellipticity and a rather constant position angle \citep[$\Delta$PA < 20\textdegree, ][]{1995wozniak1, 2007mendezdelmestre1}.

\subsection{Additional parameters}

Due to the ambiguity of radio-loudness, we used additional characteristics to estimate the contribution of the jet in these sources. We assume that sources detected at 37~GHz host a powerful, probably relativistic jet since no other known phenomena, for example, supernova remnants, are able to produce radio emission at the several hundred mJy level at such a high frequency. Seven out of our nine sources, of which three were formerly classified radio-silent, have been detected at 37~GHz. Additionally, we used the q22 parameter to estimate the jet contribution \citep{2015caccianiga1} at low radio frequencies. It is defined as the logarithm of the flux density ratio of 22~$\mu$m mid-infrared (MIR) emission and 1.4~GHz radio emission (S$_{22 \mu \text{m}}$); q22 = log (S$_{22 \mu \text{m}}$ / S$_{1.4 \text{GHz}}$), and can be used to distinguish between  sources in which the predominant source of the low frequency radio emission is the jet and sources in which the contribution of enhanced star formation is significant. It is based on the varied MIR-to-radio flux ratio in different sources; when $q22<-0.8$ the jet is presumed to be the main source of the radio emission, and when $q22>1$ the star formation may dominate the low frequency radio emission, especially in sources which have excess MIR emission, i.e., the difference between the Wide-field Infrared Survey Explorer\footnote{https://www.nasa.gov/mission\_pages/WISE/main/index.html} (WISE) magnitudes in W3$_{12 \mu \text{m}}$- and W4$_{22 \mu \text{m}}$-bands is $>$ 2.5. Sources in between, with $-0.8<q22<1$, most likely have both, the star formation and the jet, contributing to the radio emission. It should be noted that the 1.4GHz properties do not necessarily correlate with the properties at higher radio frequencies, and a source can host a powerful jet even if its low radio frequency emission is not dominated by it.

We used 1.4~GHz radio data from the Very Large Array (VLA) Faint Images of the Radio Sky at Twenty-Centimeters (FIRST) survey\footnote{http://sundog.stsci.edu/} and MIR data from WISE to calculate q22 and W3-W4 for our sample. Four out of six sources with sufficient data have $q22>1$, and the remaining two have $-0.8<q22<1$. For sources not detected at 1.4~GHz we calculated q22 assuming a flux density at the detection limit of the FIRST survey; 1~mJy. Since this is an upper limit, their q22 values are lower limits. Based on this these three radio-silent sources seem to be intermediate or star-formation dominated. Thus, according to the q22 parameter, the radio emission at 1.4~GHz is not dominated by the jet in any of these nine sources, and at least four out of nine can be assumed to exhibit enhanced star formation. Radio and infrared properties are shown in Table~\ref{tab:additional}.

\subsubsection{Black hole masses}

\begin{figure}
\centering
\includegraphics[width=0.5\textwidth]{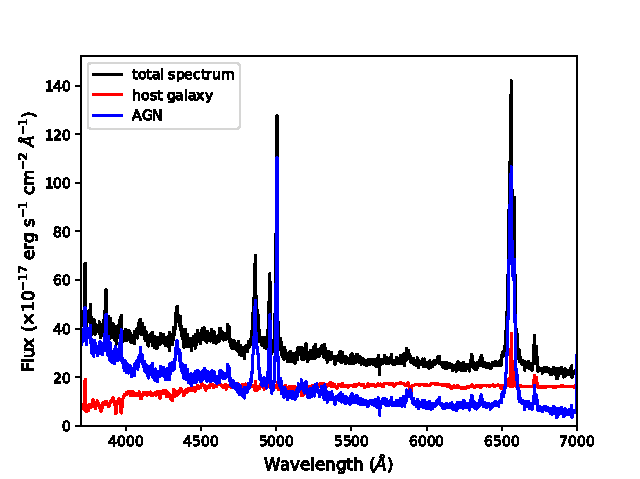}
\caption{SDSS J091313.73+365817.2. An example of the host subtraction.}
\label{fig:hostsubtraction}
\end{figure}

\begin{figure}
\centering
\includegraphics[width=0.5\textwidth]{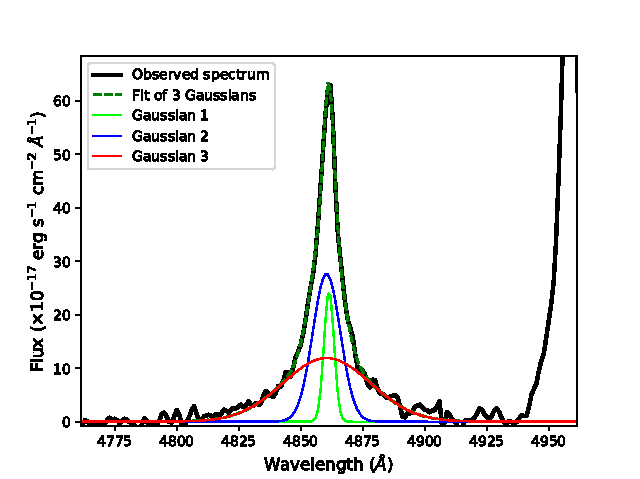}
\caption{SDSS J122749.14+321458.9. An example of the H$\beta$ fitting}
\label{fig:hbfit}
\end{figure}

The black hole masses were estimated using the second-order moment of H$\beta$ \citep{2015foschini1,2018chen1}, which is less sensitive to the orientation than FWHM(H$\beta$). We used SDSS spectrum which was corrected for redshift using, when possible, the [S~II]$\lambda$6731 line as a reference, to avoid wrong corrections due to blue outliers common among radio-loud NLS1 galaxies \citep{2016berton2}. If [S~II] was weak or absent, we used the [O~II]$\lambda$3727 line instead. The Galactic absorption correction was done using the values by \citet{2005kalberla1}. The host galaxy was subtracted from the spectra following the procedure described in \citet{2007lamura1} and \citet{2018chen1}. An example of the host subtraction is shown in Fig.~\ref{fig:hostsubtraction}. Then the power-law continuum of the AGN was subtracted. The Fe~II lines in the H$\beta$ region were subtracted using the online template\footnote{http://servo.aob.rs/FeII\_AGN/} developed by \citet{2010kovacevik1} and described in \citet{2012shapovalova1}. H$\beta$ was fitted using three Gaussian components. One was used to reproduce the narrow component and two for the broad component. Fig.~\ref{fig:hbfit} shows an example of the H$\beta$ fitting. The wavelength of the narrow component was fixed to the restframe wavelength of H$\beta$. We forced its FWHM to be smaller than or equal to the FWHM of [O~III], since the latter can be often turbulent and provide an overestimate of the narrow FWHM \citep[e.g. ][]{2015berton1}. After subtracting the narrow component, we estimated the second-order moment $\sigma$ of H$\beta$ as a velocity indicator. The $\sigma$ is less biased for BLR geometry and inclination \cite{2011peterson1}. To derive the BLR radius, we exploited the relation between the H$\beta$ flux and the BLR radius derived by \citet{2010greene1}. For the virial factor f we adopted the value of 3.85, calculated by \citet{2006collin1}. The black hole mass estimates are presented in Table~\ref{tab:additional}.

\begin{table*}
\caption[]{Radio and infrared properties of the sample, and the black hole mass estimates.}
\centering
\begin{tabular}{l l l l l l l l l l l l l}
\hline\hline
Source              & $S_{\mathrm{1.4~GHz}}$ & $S_{\mathrm{37~GHz, max}}$ & $R$  & W3-W4 & q22      & log $M_{\text{BH}}$  \\
                    & (mJy)                  & (Jy)                       &      & (mag) &          & ($M_{\odot}$) \\ \hline
J091313.73+365817.2 &  0.84                  & --                         & 1.8  & 2.55  & 1.69     & 6.9 \\ 
J111934.01+533518.7 & 17.25                  & --                         & 22.4 & 2.47  & 0.51     & 6.3 \\
J122749.14+321458.9 &  6.42                  & 0.20                       & 68.7 & 2.62  & 1.19     & 7.2 \\
J123220.11+495721.8 & --                     & 0.56                       & --   & 2.19  & 0.94$^a$ & 7.5  \\
J125635.89+500852.4 & --                     & 0.61                       & --   & 2.43  & 0.61$^a$ & 6.7 \\
J133345.47+414127.7 & 1.95                   & 0.35                       & 8.0  & 2.92  & 1.29     & 7.8 \\
J151020.06+554722.0 & --                     & 0.83                       & --   & 2.42  & 0.89$^a$ & 6.7 \\
J152205.41+393441.3 & 2.52                   & 1.43                       & 1.8  & 1.95  & 0.60     & 6.2 \\
J161259.83+421940.3 & 3.60                   & 0.64                       & 19.5 & 2.56  & 1.88     & 6.8 \\
\end{tabular}
\tablefoot{(a) Lower limit, calculated assuming 1.4~GHz flux at the detection limit of the FIRST survey, i.e., 1~mJy. }
\label{tab:additional}
\end{table*}

\section{Individual source analysis}
\label{sec:individual}

\subsection{SDSS J091313.73+365817.2} 

J091313.73+365817.2 is classified as a radio-quiet NLS1 galaxy and it has not been detected at 37~GHz. Based on parameter values q22=1.69 and W3-W4=2.55 it is expected to exhibit enhanced star formation.

\begin{figure*}[ht!]
\centering
\adjustbox{valign=t}{\begin{minipage}{0.35\textwidth}
\centering
\includegraphics[width=1\textwidth]{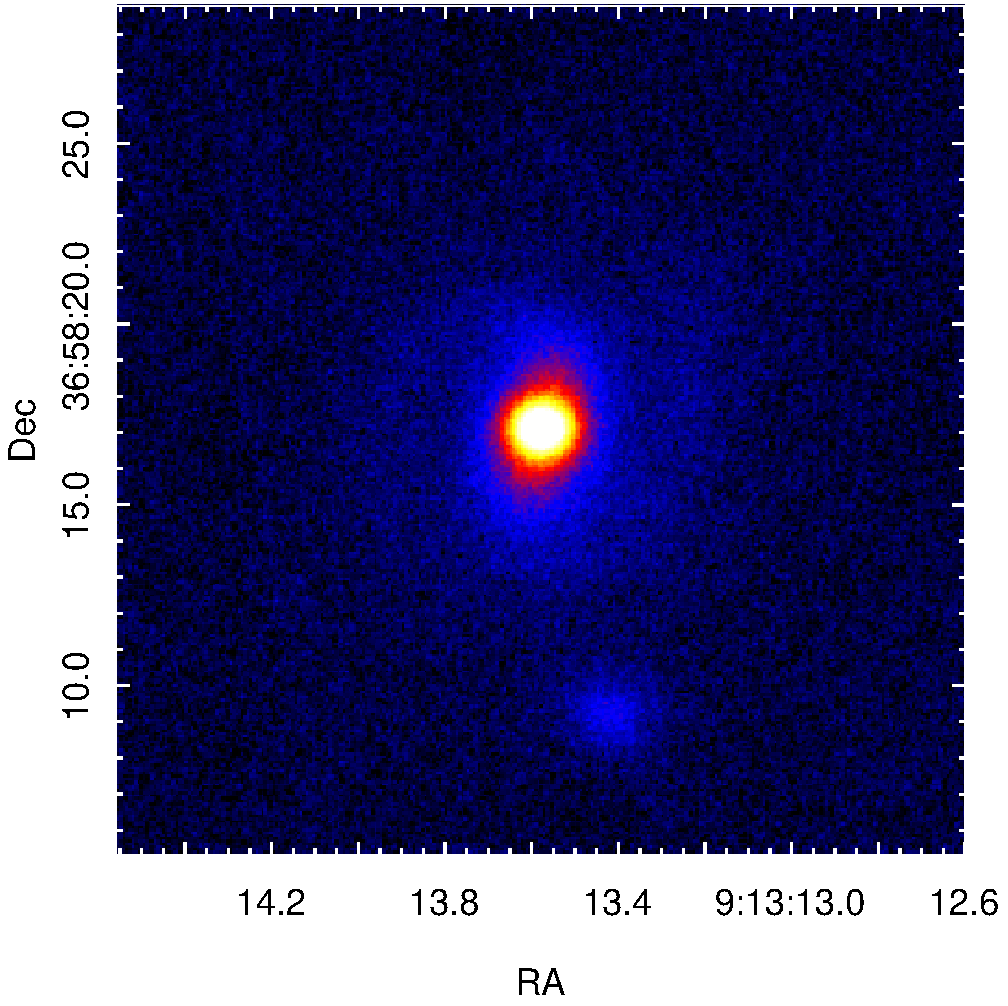}
\end{minipage}}
\adjustbox{valign=t}{\begin{minipage}{0.31\textwidth}
\centering
\includegraphics[width=0.95\textwidth]{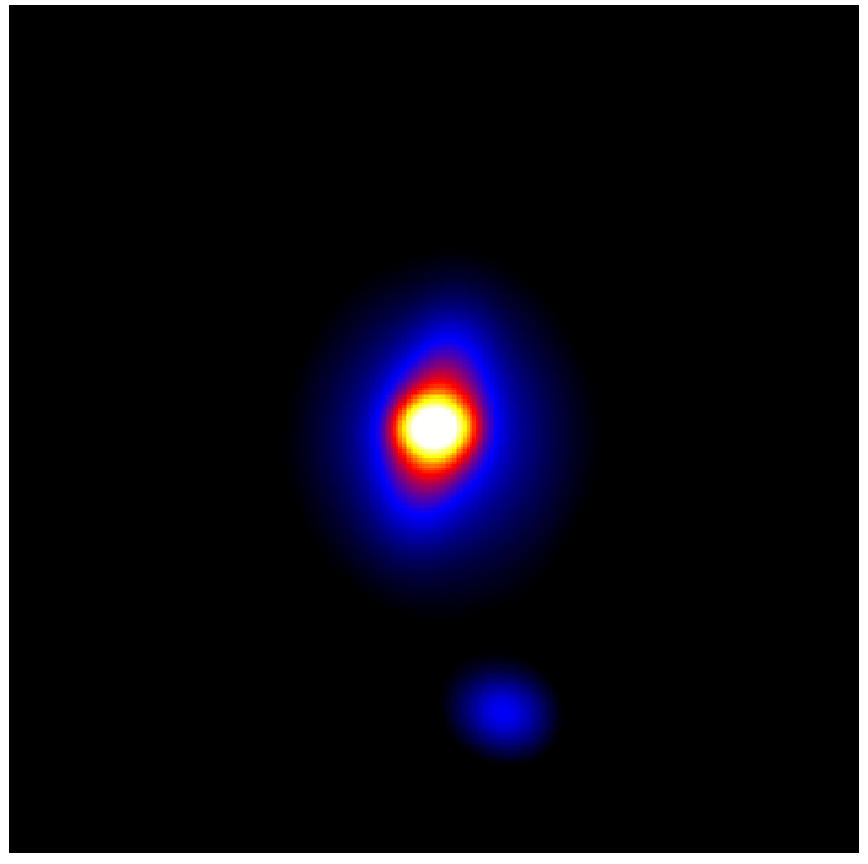}
\end{minipage}}
\adjustbox{valign=t}{\begin{minipage}{0.31\textwidth}
\centering
\includegraphics[width=0.95\textwidth]{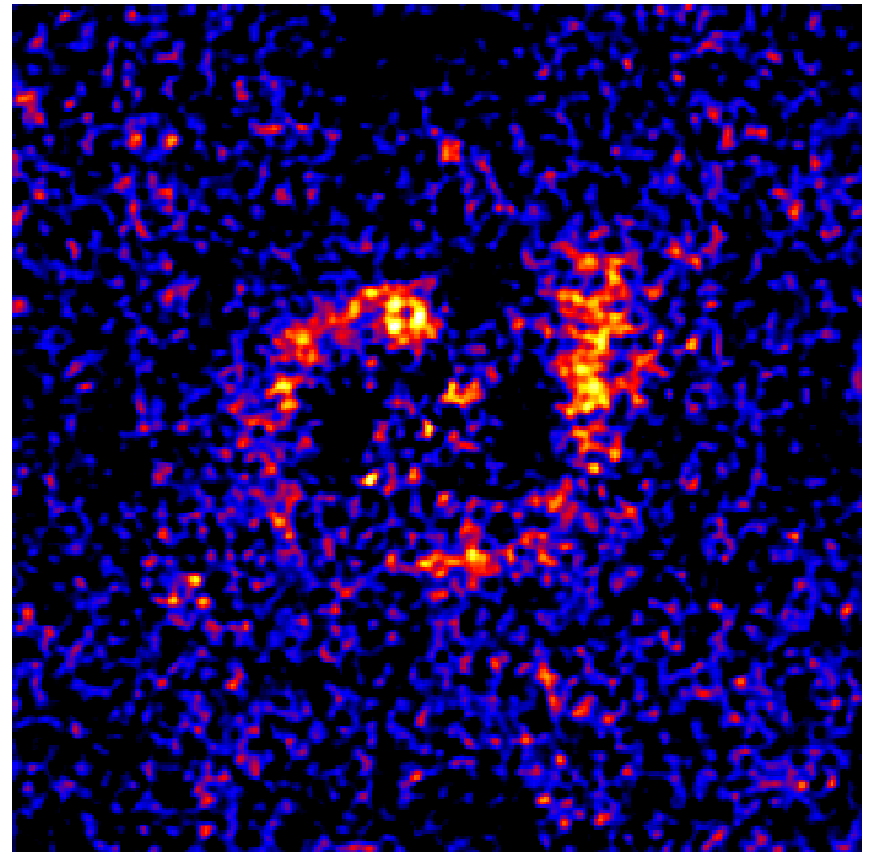}
\end{minipage}}
\hfill
    \caption{SDSS J091313.73+365817.2. The field of view is 23.4" / 44.1~kpc in all images. \emph{Left panel:} observed image, \emph{middle panel:} model image, and \emph{right panel:} residual image, smoothed over 3px.}  \label{fig:0913}
\end{figure*}

\begin{figure}
\centering
\includegraphics[width=0.5\textwidth]{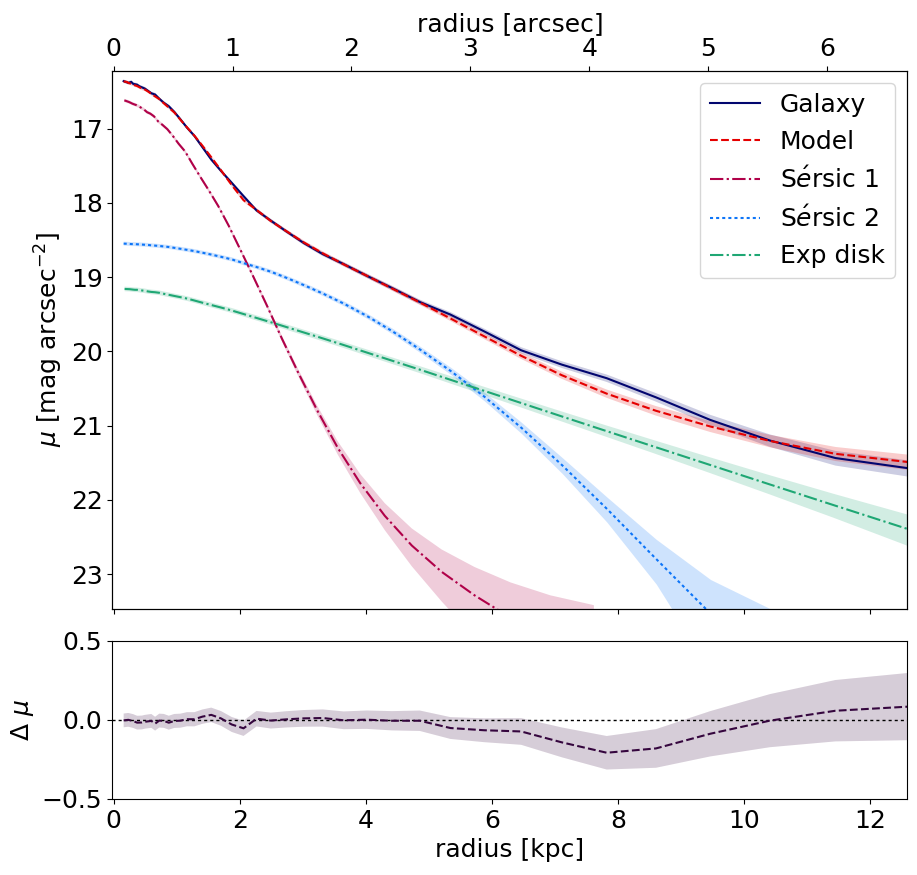}
\caption{SDSS J091313.73+365817.2. The observed and model radial surface brightness profiles. Symbols and colours are explained in the plot. The shaded area around each profile describes the associated errors.}
\label{fig:0913comps}
\end{figure}

Seeing during the PSF star observation was 0.7", but varied between $\sim$0.7"-1" during the four 900s exposures of the source itself. Visual inspection of the image clearly indicates that it is a spiral galaxy, probably barred. However, we did not try to fit the spiral arms since they are very faint. The optimal fit was achieved with two S\'{e}rsic functions, for the bulge and the bar, and an exponential function for the disk. The nearby galaxy was sufficiently fitted with a S\'{e}rsic function. The parameters of the best fit are listed in Table~\ref{tab:j0913}. The observed, model and residual images are shown in Fig.~\ref{fig:0913}, and the radial surface brightness profile in Fig.~\ref{fig:0913comps}. We tried to obtain the spectrum of the nearby galaxy using ALFOSC at NOT but were not successful; either it is too faint or has no strong emission or absorption lines. It remains unclear whether it is a satellite galaxy of J091313.73+365817.2, or a foreground or a background galaxy. Either way, with $n$=0.65 it resembles a disk galaxy. 

\begin{figure}
\centering
\includegraphics[width=0.5\textwidth]{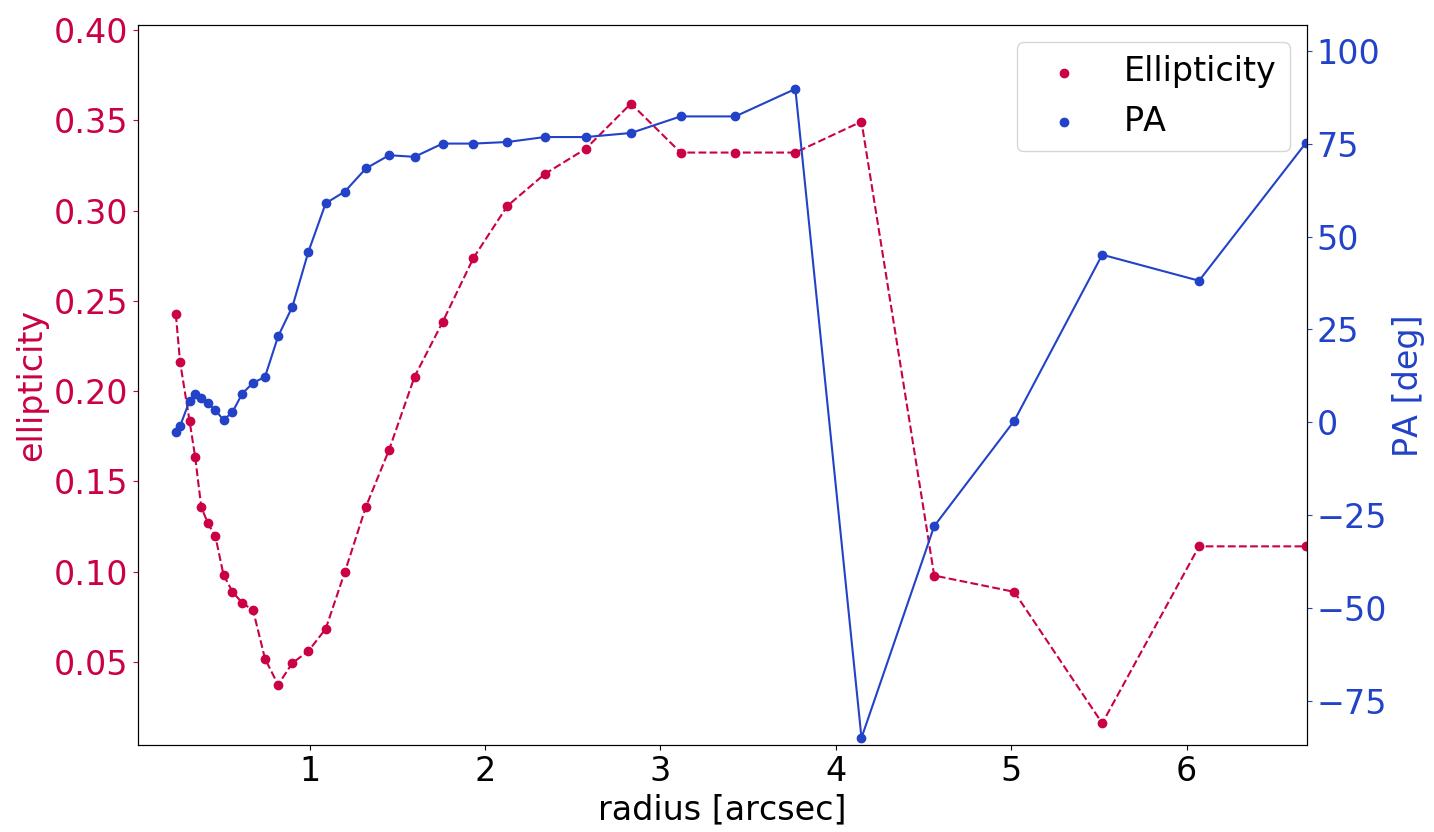}
\caption{SDSS J091313.73+365817.2. The ellipticity and the position angle plotted against the major axis of the isophote. Symbols and colours are explained in the plot.}
\label{fig:0913epa}
\end{figure}

The fit of the NLS1 source ended up being better without the PSF, and in fits where it was included it was $\sim$3 magnitudes fainter than other components, indicating that the $J$-band emission of this source is strongly dominated by the host galaxy also in the nuclear region. S\'{e}rsic 1 with $n$=0.69 and $r_e$=0.75~kpc depicts the bulge of the galaxy, which is a pseudo-bulge. The parameters of S\'{e}rsic 2 fit those of a bar with $n$=0.51, $r_e$=3.23~kpc and axial ratio of 0.40. To verify the presence of a bar we plotted ellipticity and position angle against radius. The plot is shown in Fig.~\ref{fig:0913epa}, and a region matching the description of a bar is visible between approximately 1.5--4~arcsec. The effective radius of the disk is 5.89~kpc. Since the spiral arms are not modelled they are distinct in the residual image in Fig.~\ref{fig:0913}; there are two prominent, axisymmetric arms, connected to the ends of the bar. The other arm seems to be brighter at smaller and the other at larger radii. The spiral arms also cause the radial surface brightness profile of the model to deviate from the profile of the galaxy at radii larger than $\sim$5~kpc, as seen in Fig.~\ref{fig:0913comps}.

J091313.73+365817.2 appears to be a stereotype radio-quiet NLS1, hosted by a barred spiral with a pseudo-bulge.

\renewcommand{\arraystretch}{1.5}
\begin{table*}[]
\caption[]{Best fit parameters of SDSS J091313.73+365817.2. $\chi^2_{\nu}$ = 1.12 $\substack{+0.01 \\ - 0.01}$ .}
\centering
\begin{tabular}{l l l l l l l}
\hline\hline
funct.       &  mag                              & $r_{e}$                            & $n$                              & axial                            & PA            & notes   \\
             &                                   & (kpc)                              &                                  & ratio                            & (\textdegree) &     \\ \hline
S\'{e}rsic 1 & 16.71 $\substack{+0.11 \\ -0.12}$ & 0.75 $\substack{+0.00 \\ -0.00}$   & 0.69 $\substack{+0.02 \\ -0.01}$ & 0.86 $\substack{+0.00 \\ -0.00}$ & 1.7 $\substack{+0.5 \\ -0.3}$ & bulge \\
S\'{e}rsic 2 & 17.06 $\substack{+0.11 \\ -0.10}$ & 3.23 $\substack{+0.02 \\ -0.01}$   & 0.51 $\substack{+0.00 \\ -0.02}$ & 0.40 $\substack{+0.00 \\ -0.01}$ & 77.4 $\substack{+0.4 \\ -0.3}$ & bar \\
Exp. disk    & 16.22 $\substack{+0.17 \\ -0.18}$ & 5.89 $\substack{+0.39 \\ -0.31}$   & [1.00]                           & 0.91 $\substack{+0.02 \\ -0.01}$ & -88.9 $\substack{+2.5 \\ -0.0}$ & disk \\
S\'{e}rsic 3 & 18.65 $\substack{+0.15 \\ -0.13}$ & 1.55 $\substack{+0.04 \\ -0.05}^a$ & 0.64 $\substack{+0.06 \\ -0.06}$ & 0.81 $\substack{+0.00 \\ -0.00}$ & -20.7 $\substack{+0.3 \\ -0.3}$ & nearby \\  \hline
\end{tabular}
\tablefoot{(a) Calculated assuming a redshift similar to SDSS J091313.73+365817.2.}
\label{tab:j0913}
\end{table*}

\subsection{SDSS J111934.01+533518.7}
\label{sec:j1119}

J111934.01+533518.7 is classified as a radio-loud NLS1, but it has not been detected at 37~GHz. Its q22 = 0.5 and W3-W4 = 2.5, indicating an intermediate case between jet-dominated and star-formation-dominated radio emission at lower frequencies.

\begin{figure}
\centering
\includegraphics[width=0.5\textwidth]{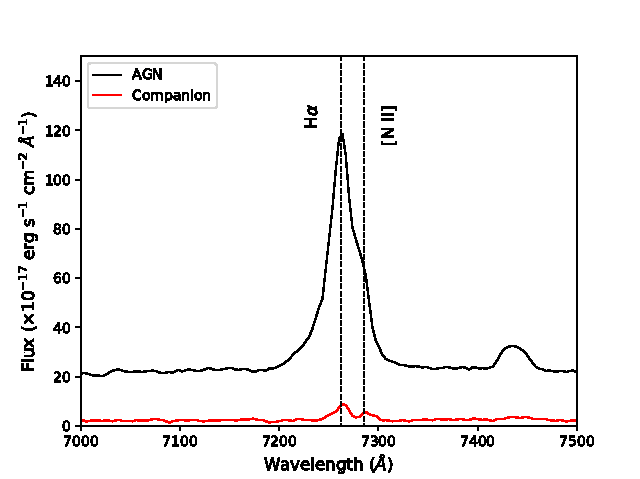}
\caption{Spectra of SDSS J111934.01+533518.7 and its companion galaxy.}
\label{fig:j1119spectra}
\end{figure}

\begin{figure*}[ht!]
\centering
\adjustbox{valign=t}{\begin{minipage}{0.35\textwidth}
\centering
\includegraphics[width=1\textwidth]{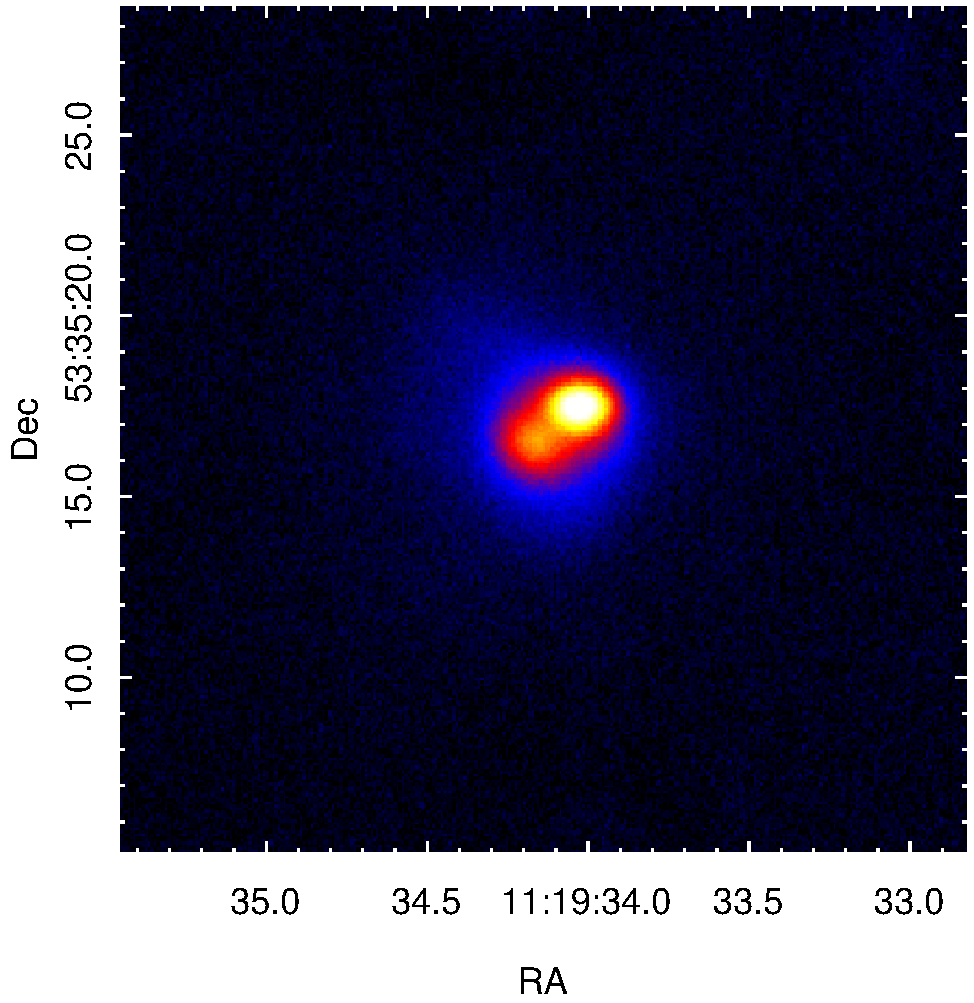}
\end{minipage}}
\adjustbox{valign=t}{\begin{minipage}{0.31\textwidth}
\centering
\includegraphics[width=1\textwidth]{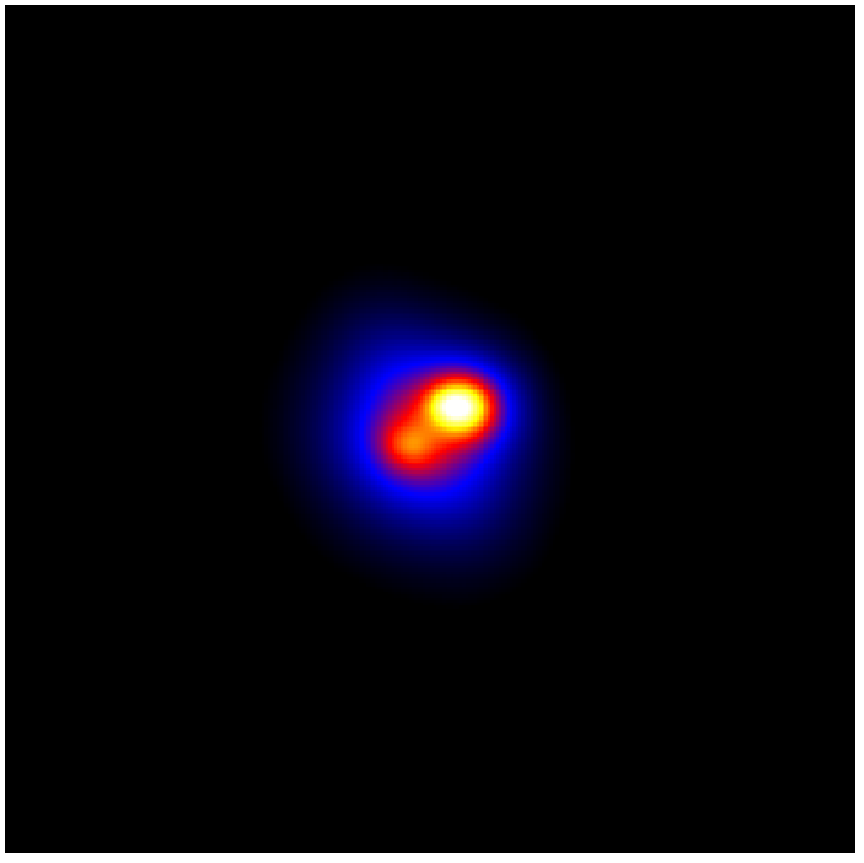}
\end{minipage}}
\adjustbox{valign=t}{\begin{minipage}{0.31\textwidth}
\centering
\includegraphics[width=1\textwidth]{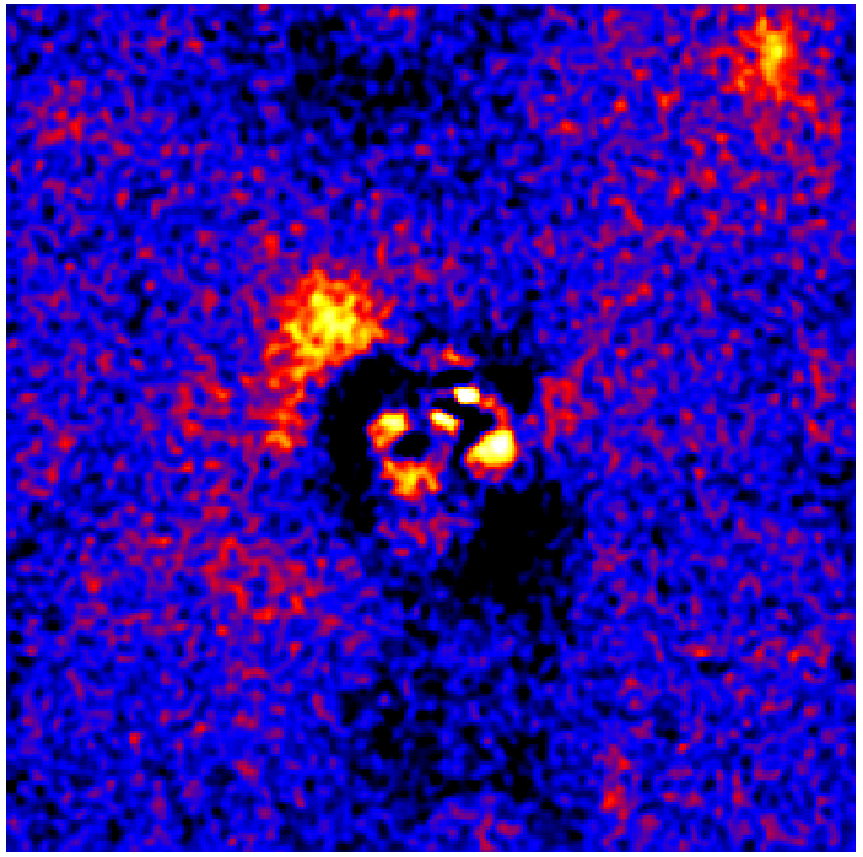}
\end{minipage}}
\hfill
    \caption{SDSS J111934.01+533518.7. The field of view is 23.4" / 44.1~kpc in all images. \emph{Left panel:} observed image. NLS1 nucleus resides in the northwest component. \emph{Middle panel:} model image, and \emph{right panel:} residual image, smoothed over 3px.} \label{fig:1119}
\end{figure*}

\begin{figure}
\centering
\includegraphics[width=0.5\textwidth]{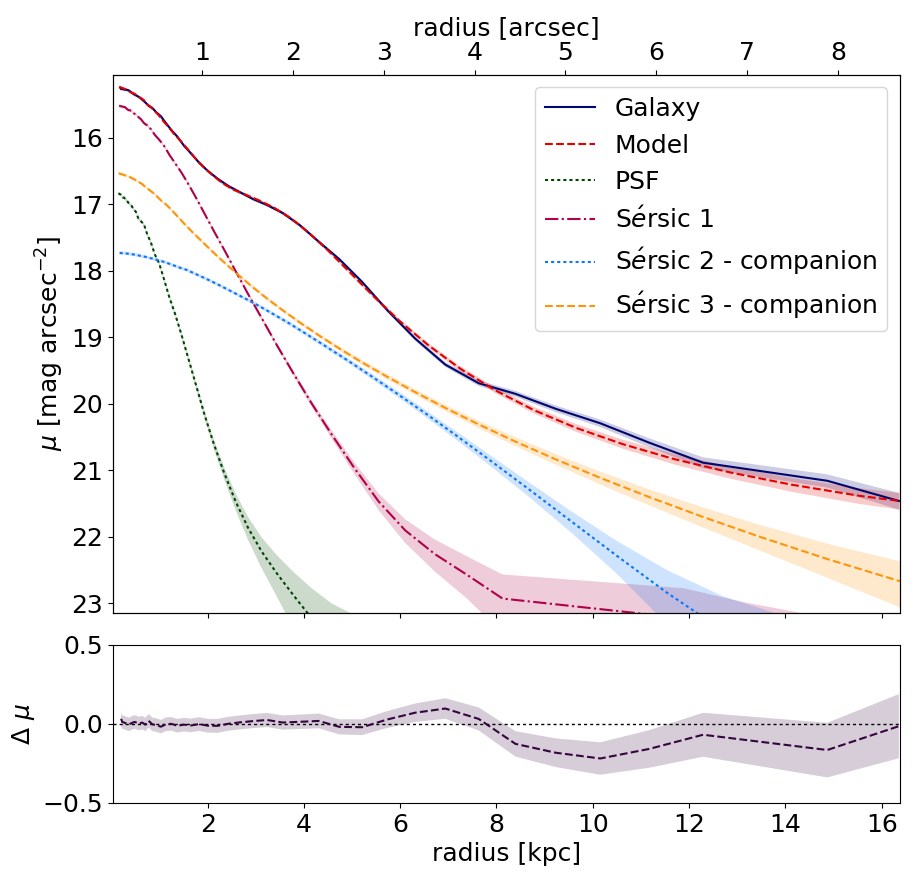}
\caption{SDSS J111934.01+533518.7. The observed and model radial surface brightness profiles. Symbols and colours are explained in the plot. The shaded area around each profile describes the associated errors.}
\label{fig:1119comps}
\end{figure}

The PSF star originally chosen for this sources had a contaminating source in the FOV, so we used the PSF star of J091313.73+365817.2 instead. The seeing during the observation was 0.7". Unfortunately the source itself has only 2 $J$-band exposures. The image shows that this is clearly an interacting source. To ascertain this we obtained spectra of the sources using ALFOSC at NOT. The observation was done on December 5, 2017, using grism $\#$4 (resolution R $\sim$360) and 1.0" slit. Total exposure time was 3600s (3 $\times$1200s), and seeing 1". The arc lamp used was HeNe, the flat lamp Halogen, and the standard star HD19445. Also an additional bias subtraction was made. The spectra of J091313.73+365817.2 and the nearby galaxy are shown in Fig.~\ref{fig:j1119spectra}. Based on emission lines visible in the spectra of both sources, H$\alpha$ and [N~II], they lie at the same redshift. The NLS1 nucleus resides in the brighter, northwest, source. We experimented with various combinations of components for this interacting pair, and the most reliable fit was achieved when the NLS1 was modelled with a PSF and a S\'{e}rsic, and the companion with two S\'{e}rsic functions. The best fit parameters are shown in Table~\ref{tab:j1119}. Other fits yielded slightly smaller $\chi^2_{\nu}$ and less residuals, but they were discarded based on the visual inspection of the subcomponents since they clearly did not correspond to actual physical components of the galaxy. Fig.~\ref{fig:1119} shows the observed, model, and residual images, and Fig.~\ref{fig:1119comps} the radial surface brightness profiles. The $r_{e}$ of the S\'{e}rsic profile of the NLS1 galaxy is 0.96~kpc and its $n$ = 1.19, compatible with a pseudo-bulge. S\'{e}rsic 3 is centered at the companion galaxy, and based on $n$ it may be an elliptical. S\'{e}rsic 2 is a bar-like component, but it is slightly off-center from the companion galaxy and fits the disturbed, elongated, structure between the two galaxies. 

The modelling of this system is likely to have been affected by its noticeably disturbed morphology. The fit leaves considerable residuals, which is understandable taking into account the interaction which perturbs the morphology. The tidal structure is a prominent feature --- also visible in the observed image --- on the northeast side of the galaxy. It is the probable cause of the model surface brightness profile deviating from the observed surface brightness profile around 8~kpc in Fig.~\ref{fig:1119comps}. 

\begin{table*}
\caption[]{Best fit parameters of SDSS J111934.01+533518.7. $\chi^2_{\nu}$ = 1.18 $\substack{+0.00 \\ -0.00}$.}
\centering
\begin{tabular}{l l l l l l l}
\hline\hline
funct.       &  mag                              & $r_{e}$                          & $n$                              & axial & PA            & notes   \\
             &                                   & (kpc)                            &                                  & ratio & (\textdegree) &  \\ \hline
PSF          & 17.66 $\substack{+0.81 \\ -0.19}$ &                                  &                                  &       &               &   \\
S\'{e}rsic 1 & 15.34 $\substack{+0.13 \\ -0.22}$ & 0.96 $\substack{+0.03 \\ -0.00}$ & 1.19 $\substack{+0.18 \\ -0.01}$ & 0.90 $\substack{+0.01 \\ -0.05}$ & -16.5  $\substack{+7.4 \\ -7.0}$ & bulge \\
S\'{e}rsic 2 & 16.09 $\substack{+1.36 \\ -0.42}$ & 3.46 $\substack{+1.81 \\ -0.29}$ & 0.81 $\substack{+0.05 \\ -0.75}$ & 0.51 $\substack{+0.06 \\ -0.25}$ & -61.4  $\substack{+16.8 \\ -2.8}$ & companion \\  
S\'{e}rsic 3 & 15.05 $\substack{+0.00 \\ -0.19}$ & 3.79 $\substack{+1.78 \\ -0.68}$ & 2.46 $\substack{+0.92 \\ -0.79}$ & 0.85 $\substack{+0.00 \\ -0.07}$ & -34.5 $\substack{+11.6 \\ -14.22}$ & companion \\  \hline   
\end{tabular}
\label{tab:j1119}
\end{table*}

\subsection{SDSS J151020.06+554722.0} 

J151020.06+554722.0 was not detected in FIRST survey and was therefore previously classified as radio-silent. However, it has been frequently detected at 37~GHz, with a maximum flux density of 0.83~Jy, indicating the presence of a relativistic jet \citep{2018lahteenmaki1}. Assuming a 1.4~GHz flux density at the detection limit of FIRST, 1~mJy, the star formation tracers give q22 = 0.89 and W3-W4 = 2.42. These are lower limits and imply that the low radio frequency emission of this source is not jet-dominated.

\begin{figure}
\centering
\includegraphics[width=0.5\textwidth]{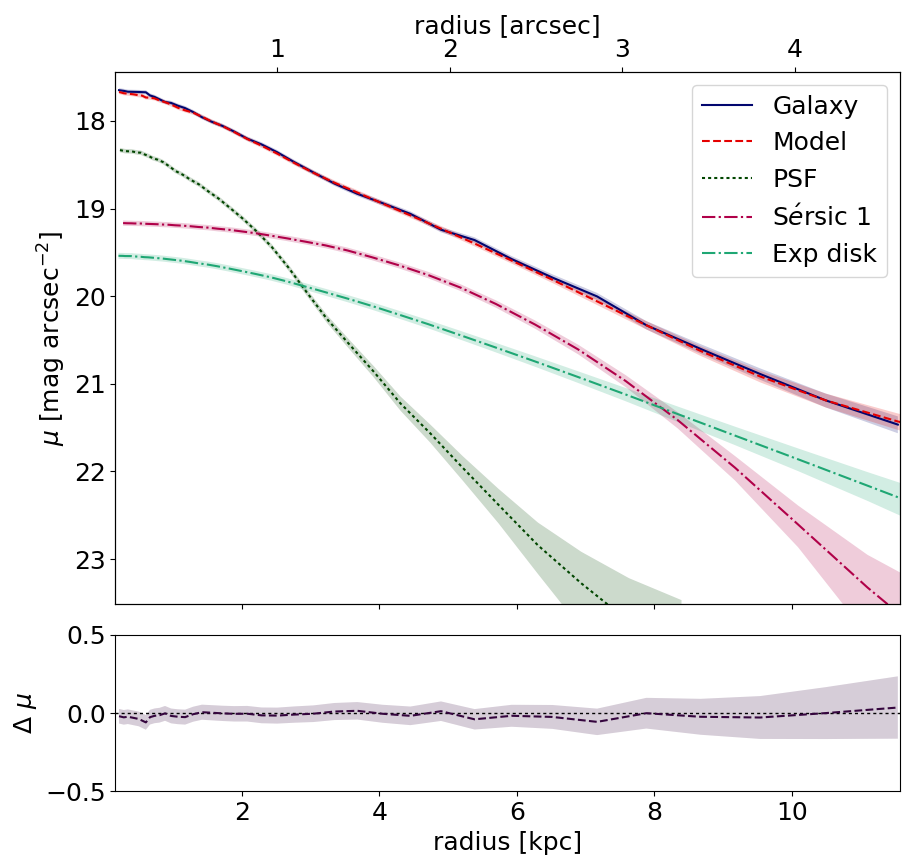}
\caption{SDSS J151020.06+554722.0. The observed and model radial surface brightness profiles. Symbols and colours are explained in the plot. The shaded area around each profile describes the associated errors.}
\label{fig:1510comps}
\end{figure}

\begin{figure}
\centering
\includegraphics[width=0.5\textwidth]{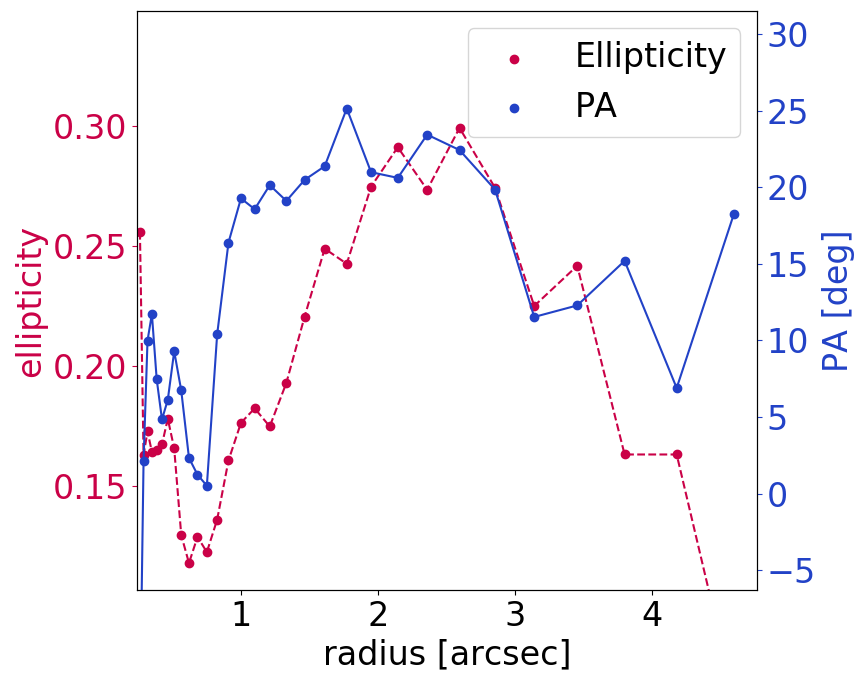}
\caption{SDSS J151020.06+554722.0. The ellipticity and the position angle plotted against the major axis of the isophote. Symbols and colours are explained in the plot.}
\label{fig:1510epa}
\end{figure}

\begin{figure*}[ht!]
\centering
\adjustbox{valign=t}{\begin{minipage}{0.34\textwidth}
\centering
\includegraphics[width=1\textwidth]{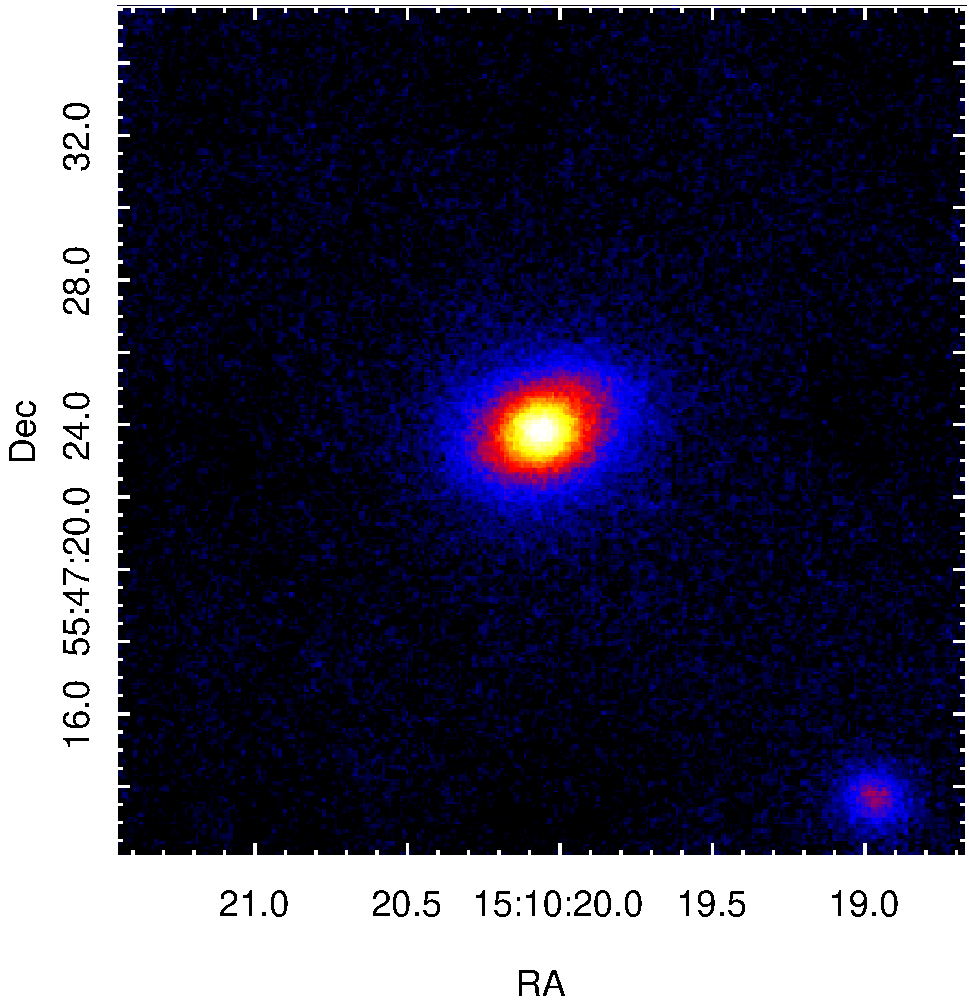}
\end{minipage}}
\adjustbox{valign=t}{\begin{minipage}{0.32\textwidth}
\centering
\includegraphics[width=0.95\textwidth]{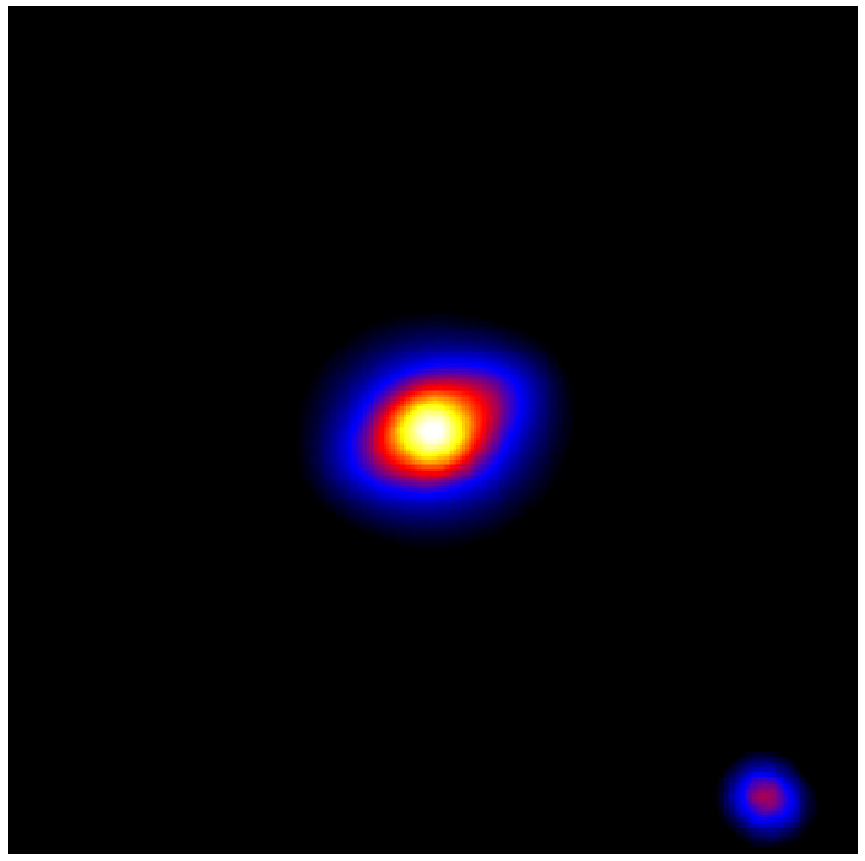}
\end{minipage}}
\adjustbox{valign=t}{\begin{minipage}{0.32\textwidth}
\centering
\includegraphics[width=0.95\textwidth]{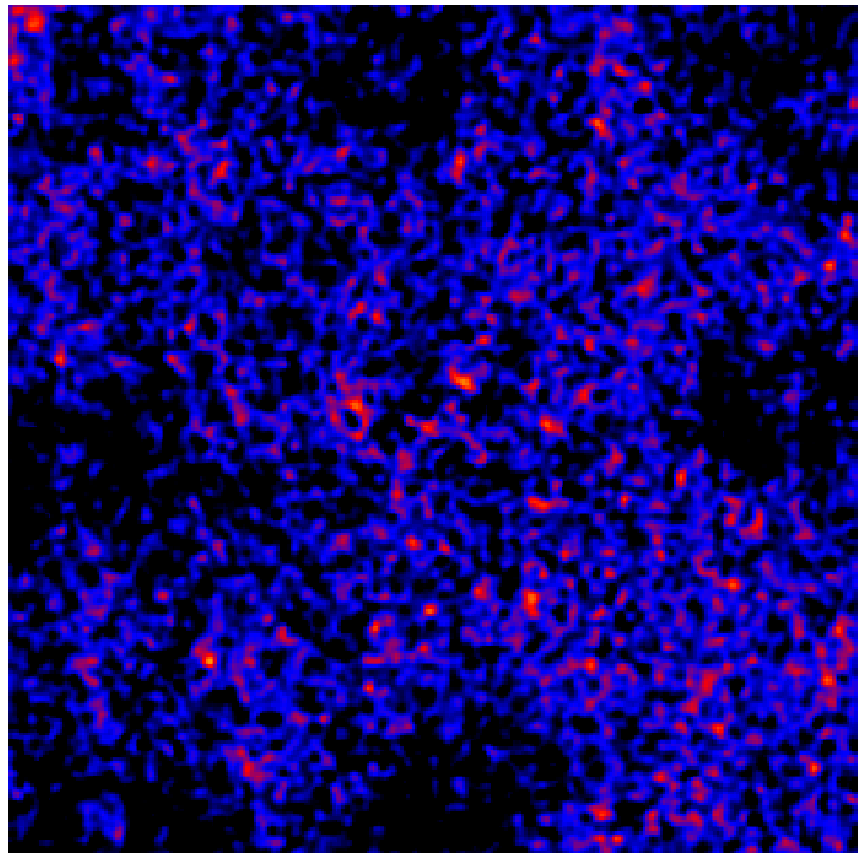}
\end{minipage}}
\hfill
    \caption{SDSS J151020.06+554722.0. The field of view is 23.4" / 58.8~kpc in all images. \emph{Left panel:} observed image, \emph{middle panel:} model image, and \emph{right panel:} residual image, smoothed over 3px.} \label{fig:1510}
\end{figure*}

The FWHM of the PSF star is 1.3", so the seeing was not optimal during this observation. We have in total four exposures, of which two unfortunately have even worse seeing ( 1.7"). In addition to J151020.06+554722.0 we also fitted a nearby source, which was nicely modelled with a PSF. The best fit for J151020.06+554722.0 was achieved with a PSF, a S\'{e}rsic and en exponential disk functions. The parameters are listed in Table~\ref{tab:j1510}, the observed, model, and residual images are shown in Fig.~\ref{fig:1510}, and Fig.~\ref{fig:1510comps} shows the radial surface brightness profiles. The S\'{e}rsic function, with $n$=0.32 and an axial ratio of 0.30, accounts for the bar with $r_e$=4.18~kpc. The $\epsilon$ and PA versus radius plot is presented in Fig.~\ref{fig:1510epa}, and it shows a region that can be interpreted as a bar around 1--3~arcsec. The effective radius of the disk is 5.31~kpc. The radial surface brightness profiles of the galaxy and the model agree well and almost no residuals are left in Fig.~\ref{fig:1510}. Thus, despite the non-optimal data quality, it is safe to assume that the host galaxy of J151020.06+554722.0 is a disk-like galaxy.

\begin{table*}
\caption[]{Best fit parameters of SDSS J151020.06+554722.0. $\chi^2_{\nu}$ = 1.12$\substack{+0.01 \\ -0.00}$.}
\centering
\begin{tabular}{l l l l l l l l}
\hline\hline
function   &  Mag                              & $r_{e}$                          & $n$                              & axial                            & PA             & notes      \\
           &                                   & (kpc)                            &                                  & ratio                            & (\textdegree)  &            \\ \hline
PSF        & 17.83 $\substack{+0.13 \\ -0.12}$ &                                  &                                  &                                  &                &            \\
S\'{e}rsic & 17.44 $\substack{+0.28 \\ -0.35}$ & 4.18 $\substack{+0.14 \\ -0.10}$ & 0.32 $\substack{+0.07 \\ -0.07}$ & 0.30 $\substack{+0.07 \\ -0.04}$ & 27.7 $\substack{+0.9 \\ -1.1}$ & bar  \\
Expdisk    & 17.04 $\substack{+0.15 \\ -0.12}$ & 5.31 $\substack{+1.76 \\ -0.69}$ & [1]                              & 0.86 $\substack{+0.00 \\ -0.03}$ & -8.6 $\substack{+9.5 \\ -22.1}$ & disk \\
PSF        & 18.71 $\substack{+0.12 \\ -0.13}$ &               &      &       &            & nearby source \\  \hline
\end{tabular}
\label{tab:j1510}
\end{table*}

\subsection{SDSS J152205.41+393441.3}

J152205.41+393441.3 is a radio-quiet NLS1 galaxy, detected at 37~GHz at an unusually high level with a maximum flux density of 1.43~Jy. Based on the values q22 = 0.6 and W3-W4 = 1.95 it does not have particularly high star formation activity, but it should not be jet-dominated either at low radio frequencies.

\begin{figure}
\centering
\includegraphics[width=0.5\textwidth]{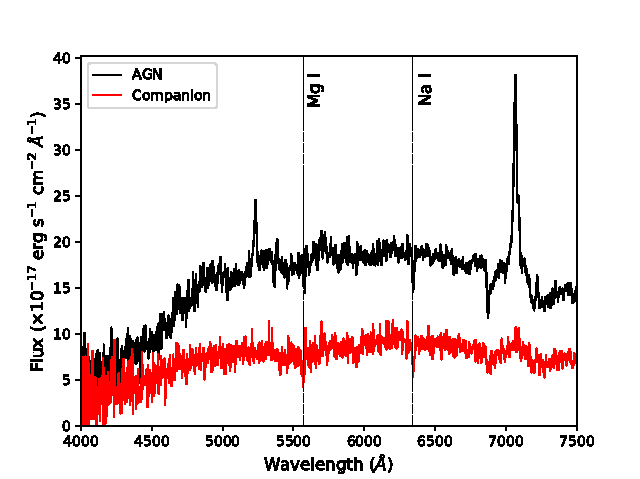}
\caption{Spectra of SDSS J152205.41+393441.3 and its companion galaxy.}
\label{fig:j1522spectra}
\end{figure}

\begin{figure*}[ht!]
\centering
\adjustbox{valign=t}{\begin{minipage}{0.35\textwidth}
\centering
\includegraphics[width=1\textwidth]{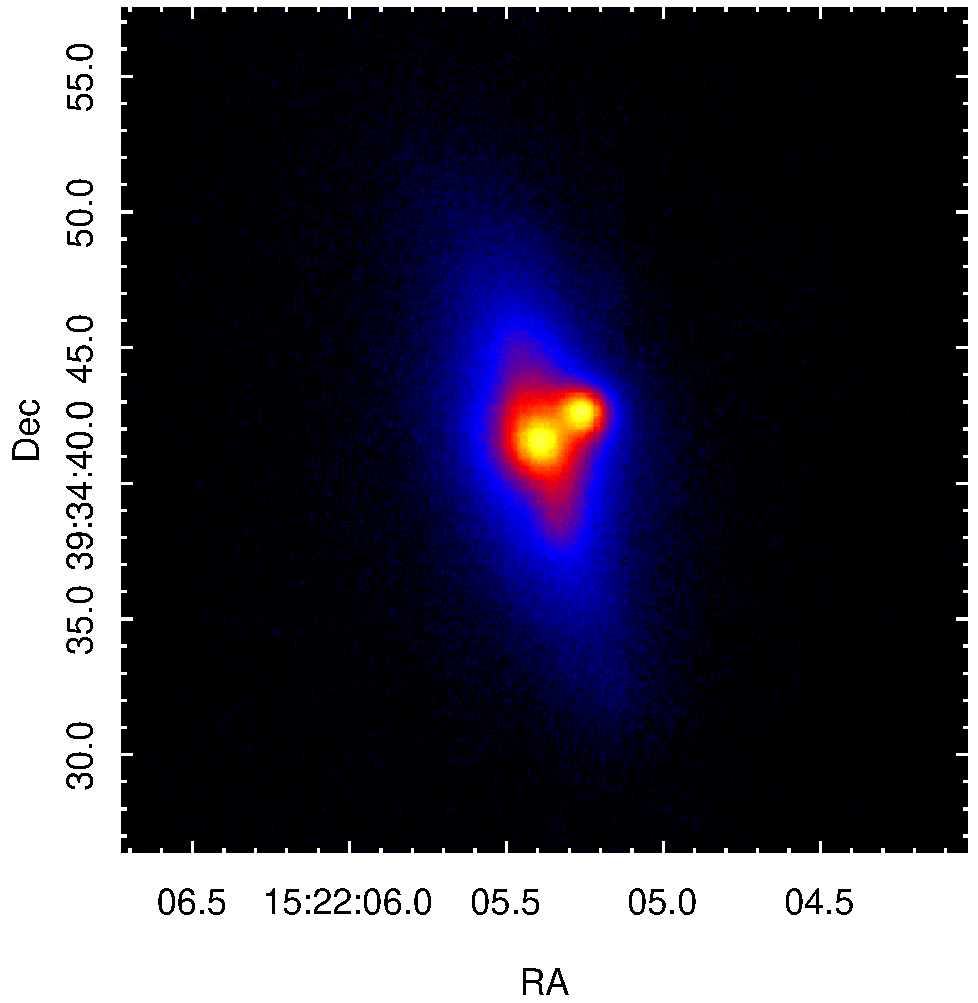}
\end{minipage}}
\adjustbox{valign=t}{\begin{minipage}{0.31\textwidth}
\centering
\includegraphics[width=1\textwidth]{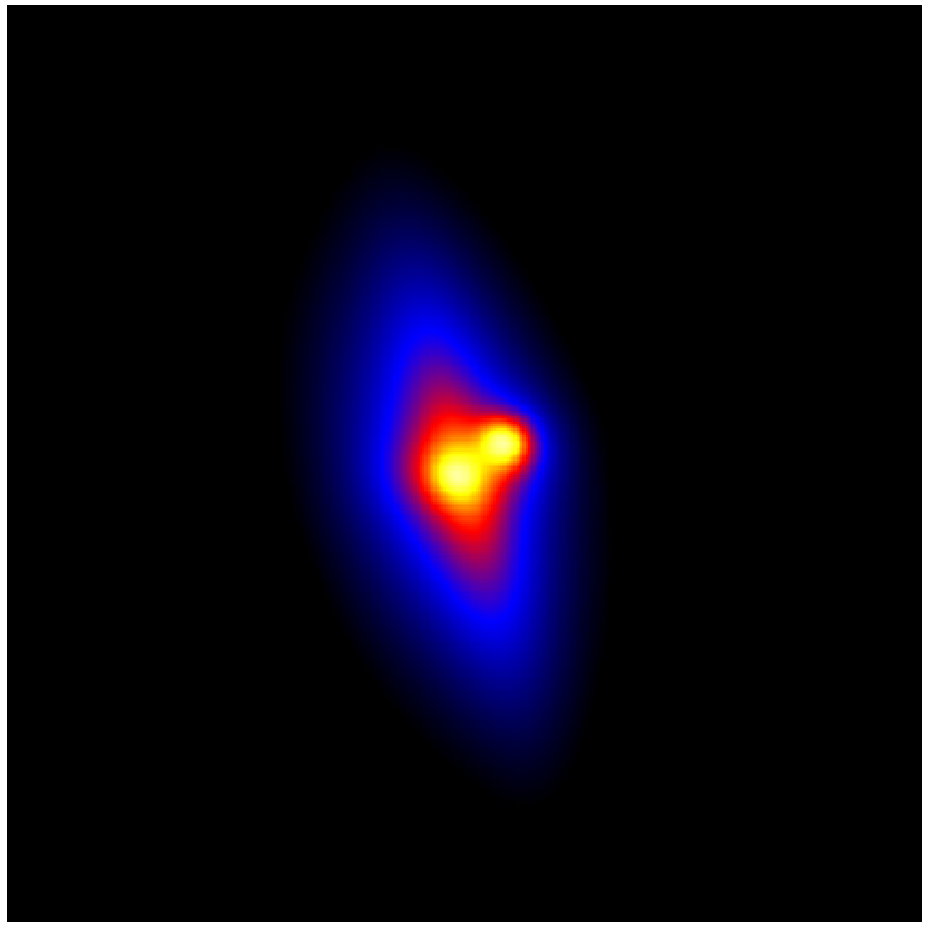}
\end{minipage}}
\adjustbox{valign=t}{\begin{minipage}{0.31\textwidth}
\centering
\includegraphics[width=1\textwidth]{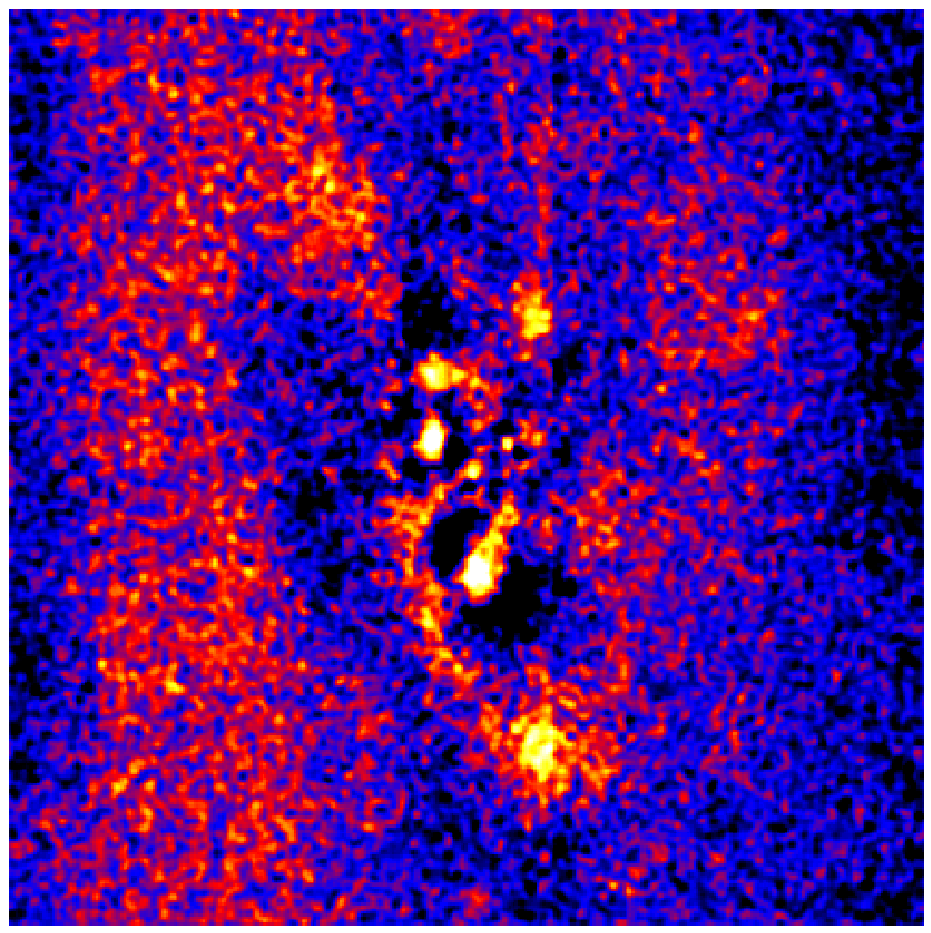}
\end{minipage}}
\hfill
    \caption{SDSS J152205.41+393441.3. The field of view is 31.2" / 43.5~kpc in all images.  \emph{Left panel:} observed image. NLS1 nucleus resides in the eastern galaxy. \emph{Middle panel:} model image, and \emph{right panel:} residual image, smoothed over 3px.} \label{fig:1522}
\end{figure*}

\begin{figure}
\centering
\includegraphics[width=0.5\textwidth]{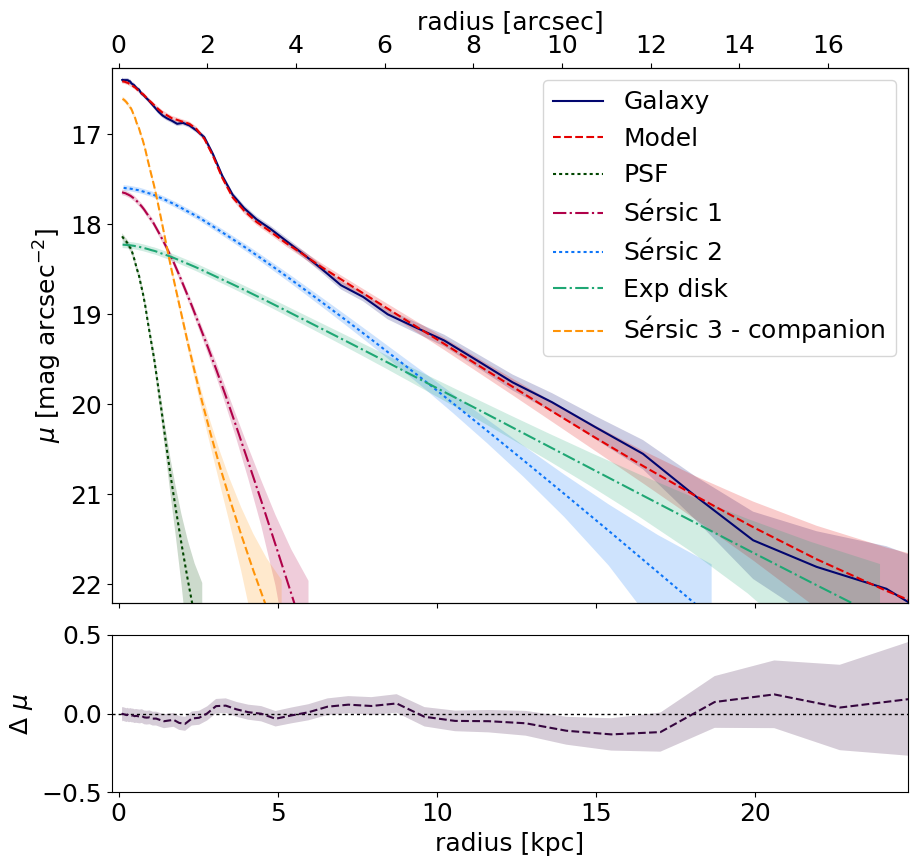}
\caption{SDSS J152205.41+393441.3. The observed and model radial surface brightness profiles. Symbols and colours are explained in the plot. The shaded area around each profile describes the associated errors.}
\label{fig:1522comps}
\end{figure}

\begin{figure}
\centering
\includegraphics[width=0.5\textwidth]{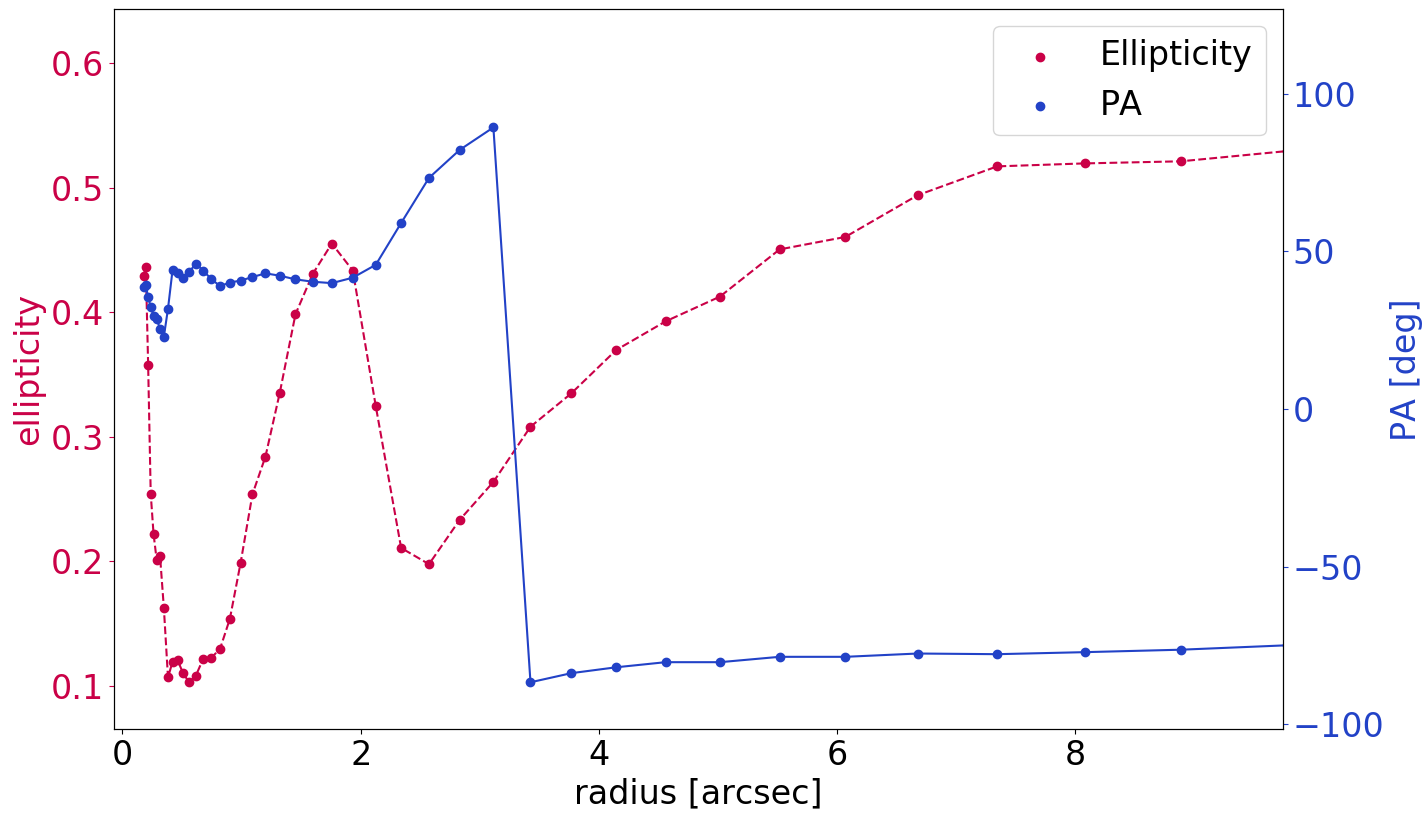}
\caption{SDSS J152205.41+393441.3. The ellipticity and the position angle plotted against the major axis of the isophote. Symbols and colours are explained in the plot.}
\label{fig:1522epa}
\end{figure}

We have 3600s of exposure time for this source, but the seeing was rather poor, $\sim$1.2". However, it is apparent from the original image in Fig.~\ref{fig:1522} that this is an interacting source. 
To ascertain this we obtained the spectra of both sources with the Asiago 1.82m Copernico telescope. The observation was done on October 16, 2017. The exposure time was 3600s, and the seeing 1". We 
used grism $\#$4 (resolution R $\sim$486). Arc lamps used were HgCd and Neon, and the standard star was HR7596. The spectra are shown in Fig.~\ref{fig:j1522spectra}. The absorption lines of the 
companion galaxy are at the same redshift as the emission lines of the NLS1 galaxy. The NLS1 nucleus lies in the more extended, eastern, galaxy, resembling a barred spiral in visual inspection. 
This is a complicated source and almost equally good fits could be achieved with many combinations of functions. In addition to $\chi^2_{\nu}$ we inspected the subcomponents of the fits carefully 
and compared the radial surface brightness profiles of the models to the observed one. Based on these criteria the optimal, and also realistic, fit for the NLS1 galaxy was achieved with a PSF, 
S\'{e}rsic functions for the bulge and the bar, and an exponential function for the disk. The companion was best modelled with one S\'{e}rsic function. The model parameters are given in 
Table~\ref{tab:j1522v2}. Fig.~\ref{fig:1522} shows the observed, model, and residual images, and the radial surface brightness profiles are shown in Fig.~\ref{fig:1522comps}. The model parameters 
confirm that J152205.41+393441.3 is a late-type galaxy. The bulge component, S\'{e}rsic 1, has $n$=0.84 suggesting that it is a pseudo-bulge with an effective radius of 1.62~kpc. S\'{e}rsic 2 
component clearly resembles a bar with $n$=0.96, $r_e$=6.14~kpc and an axial ratio of 0.24. To study the bar more closely we plotted $\epsilon$/PA versus radius in Fig.~\ref{fig:1522epa}. A region 
fitting the description of a bar is seen between 0.75--2.5~arcsec. The effective radius of the disk is 9.24~kpc. The bulge model is slightly extended towards the companion galaxy, explaining the 
rather small axial ratio and PA different compared to the bar and disk components. Based on the S\'{e}rsic 3 profile parameters, $n$=2.83 and $r_e$=0.51~kpc, the companion galaxy is likely 
to be a very small or a dwarf elliptical galaxy. All of the fit parameters are likely to be affected to some extent by the interaction. Considerable residuals are visible in the residual 
image in Fig.~\ref{fig:1522}, owing to the disturbed morphology of the system. Large-scale residuals with S/N = 4 and possibly resembling tidal structures are seen on both sides of the galaxy.

\begin{table*}
\caption[]{Best fit parameters of SDSS J152205.41+393441.3. $\chi^2_{\nu}$= 1.05 $\substack{+0.21 \\ -0.13}$.}
\centering
\begin{tabular}{l l l l l l l l}
\hline\hline
funct        & mag                               & $r_{e}$                          & $n$                              & axial                            & PA               & notes   \\
             &                                   & (kpc)                            &                                  & ratio                            & (\textdegree)    &     \\ \hline
PSF          & 18.24 $\substack{+0.10 \\ -0.10}$ &                                  &                                  &                                  &                  &     \\
S\'{e}rsic 1 & 16.38 $\substack{+0.18 \\ -0.17}$ & 1.62 $\substack{+0.12 \\ -0.09}$ & 0.84 $\substack{+0.08 \\ -0.01}$ & 0.63 $\substack{+0.00 \\ -0.01}$ & -0.5 $\substack{+0.7 \\ -0.2}$ & bulge \\ 
S\'{e}rsic 2 & 14.96 $\substack{+0.48 \\ -0.38}$ & 6.14 $\substack{+0.12 \\ -0.02}$ & 0.96 $\substack{+0.00 \\ -0.02}$ & 0.24 $\substack{+0.04 \\ -0.04}$ & -76.5 $\substack{+0.2 \\ -0.3}$ & bar  \\
Expdisk      & 14.30 $\substack{+0.10 \\ -0.16}$ & 9.24 $\substack{+2.75 \\ -1.55}$ & [1]                              & 0.53 $\substack{+0.03 \\ -0.03}$ & -75.9 $\substack{+0.0 \\ -0.5}$ & disk  \\
S\'{e}rsic 3 & 16.18 $\substack{+0.13 \\ -0.10}$ & 0.51 $\substack{+0.00 \\ -0.01}$ & 2.83 $\substack{+0.00 \\ -0.20}$ & 0.93 $\substack{+0.00 \\ -0.03}$ & -12.4 $\substack{+14.0 \\ -1.8}$ & companion \\  \hline
\end{tabular}
\label{tab:j1522v2}
\end{table*}

\subsection{SDSS J161259.83+421940.3}

J161259.83+421940.3 is a radio-loud flat-spectrum NLS1 galaxy, also detected at 37~GHz at a maximum flux density level of 0.64~Jy. Based on the infrared characteristics, q22 = 1.9 and W3-W4 = 2.6, this source exhibits enhanced star formation; \citet{2015caccianiga1} estimated the SFR to be 224 $M_{\odot}$ yr$^{-1}$, which is very high and indicates starburst activity. However, the flat radio spectrum and the 37~GHz detection suggests that also a relativistic jet is present. The kpc-scale radio morphology of J161259.83+421940.3 remains unclear due to a nearby contaminating source in JVLA observations \citep{2018berton1}.

\begin{figure}
\centering
\includegraphics[width=0.35\textwidth]{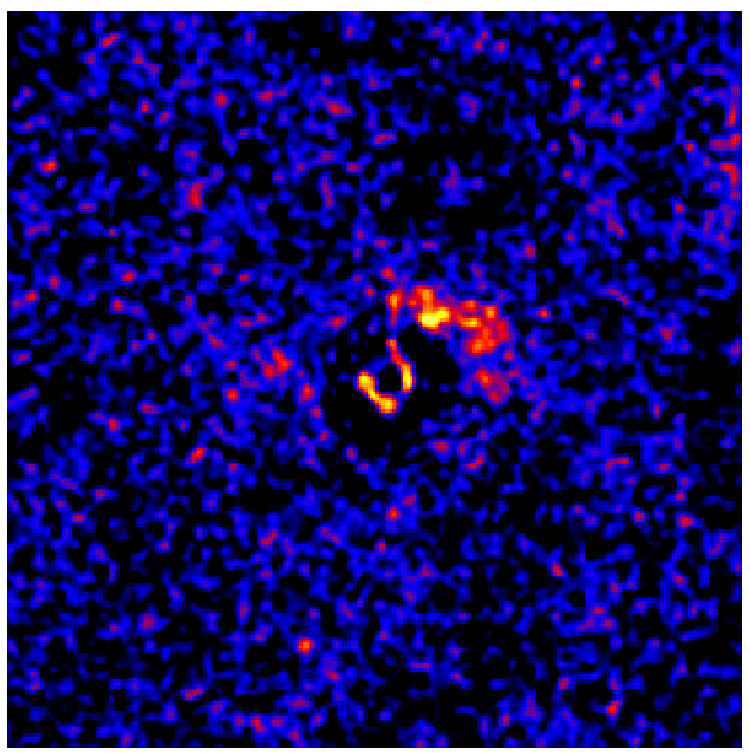}
\caption{SDSS J161259.83+421940.3. The residuals after fitting with a PSF and one S\'{e}rsic function. The field of view is 20.3" / 72.5~kpc.}
\label{fig:1612nonsym}
\end{figure}

\begin{figure}
\centering
\includegraphics[width=0.5\textwidth]{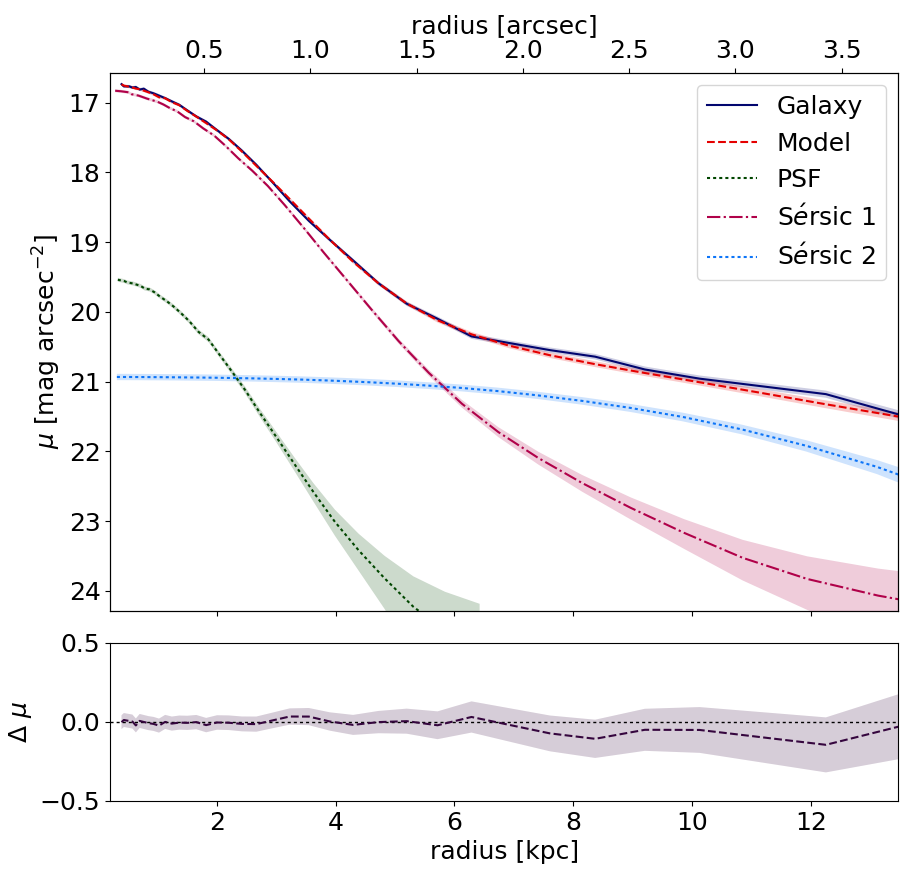}
\caption{SDSS J161259.83+421940.3. The observed and model radial surface brightness profiles. Symbols and colours are explained in the plot.The shaded area around each profile describes the associated errors.}
\label{fig:1612comps}
\end{figure}

\begin{figure*}[ht!]
\centering
\adjustbox{valign=t}{\begin{minipage}{0.36\textwidth}
\centering
\includegraphics[width=1\textwidth]{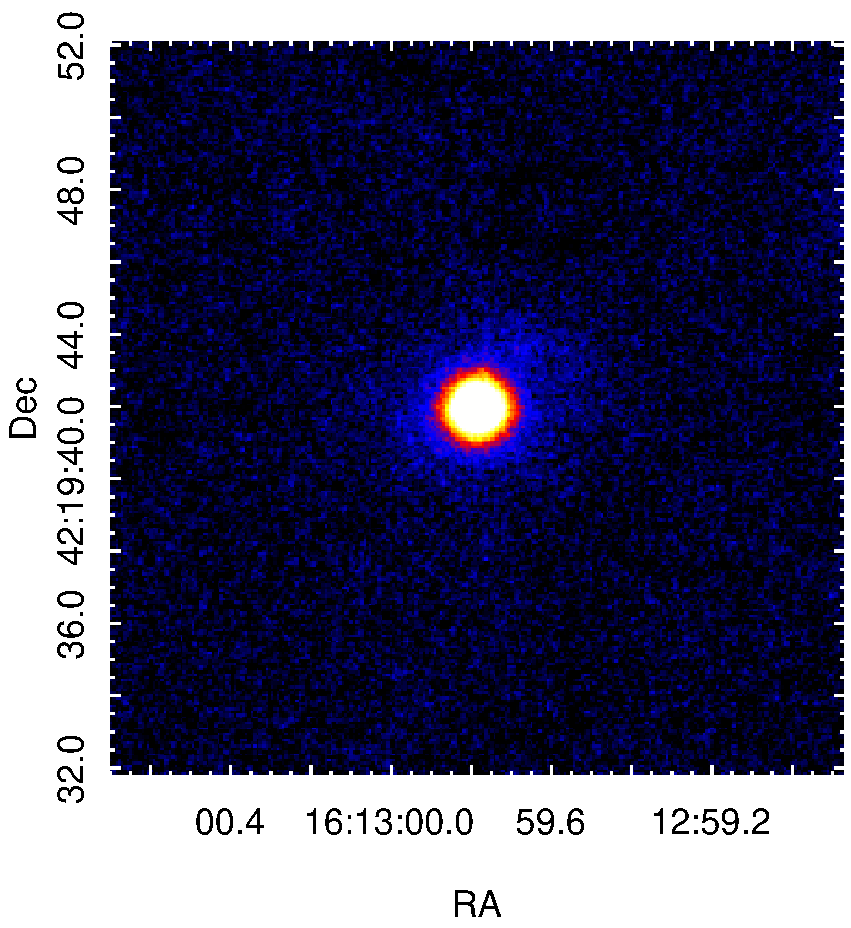}
\end{minipage}}
\adjustbox{valign=t}{\begin{minipage}{0.31\textwidth}
\centering
\includegraphics[width=1\textwidth]{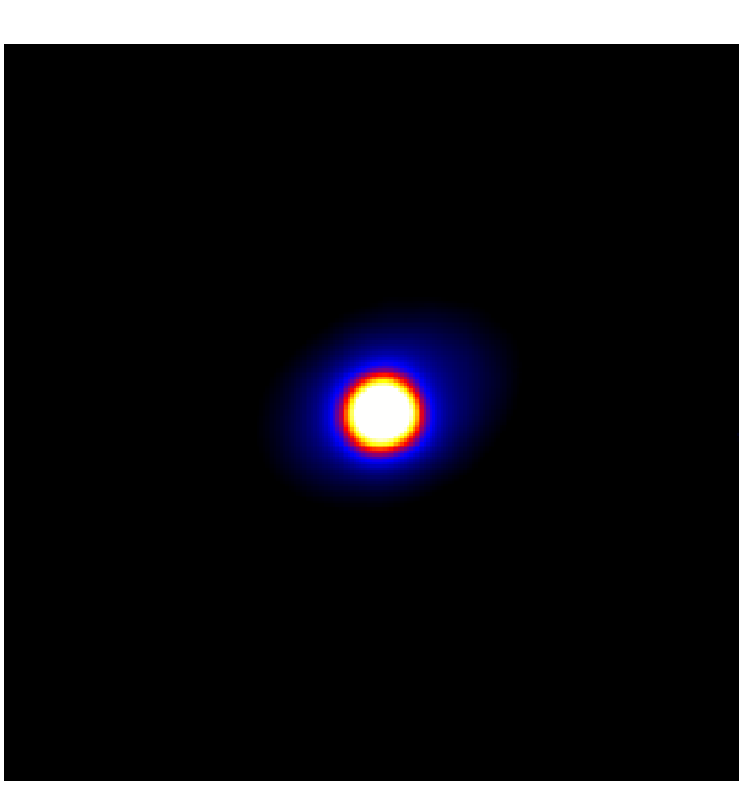}
\end{minipage}}
\adjustbox{valign=t}{\begin{minipage}{0.31\textwidth}
\centering
\includegraphics[width=1\textwidth]{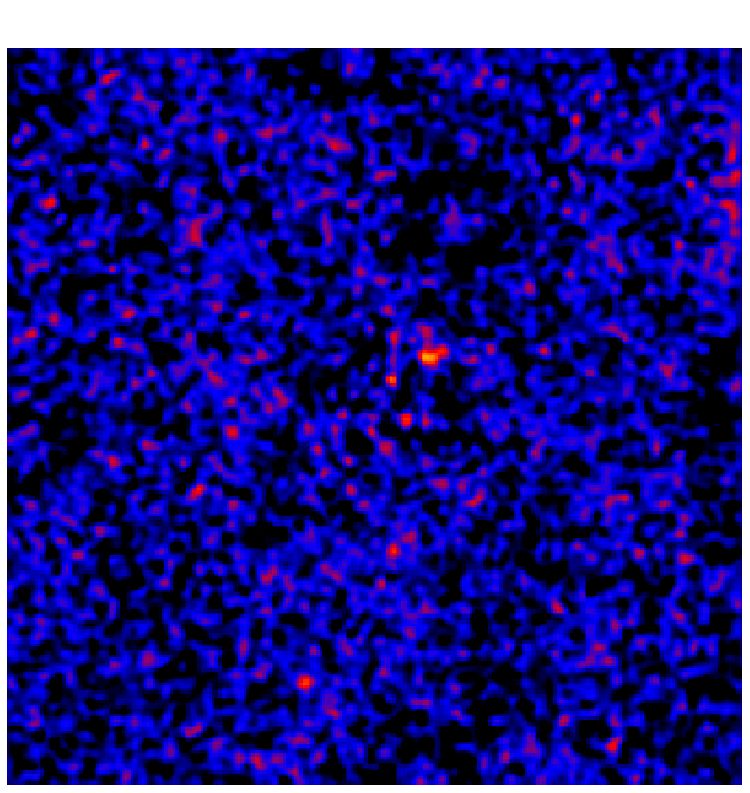}
\end{minipage}}
\hfill
    \caption{SDSS J161259.83+421940.3. The field of view is 20.3" / 72.5~kpc in all images.  \emph{Left panel:} observed image, \emph{middle panel:} model image, and \emph{right panel:} residual image, smoothed over 3px.} \label{fig:1612}
\end{figure*}

\begin{figure}
\centering
\includegraphics[width=0.5\textwidth]{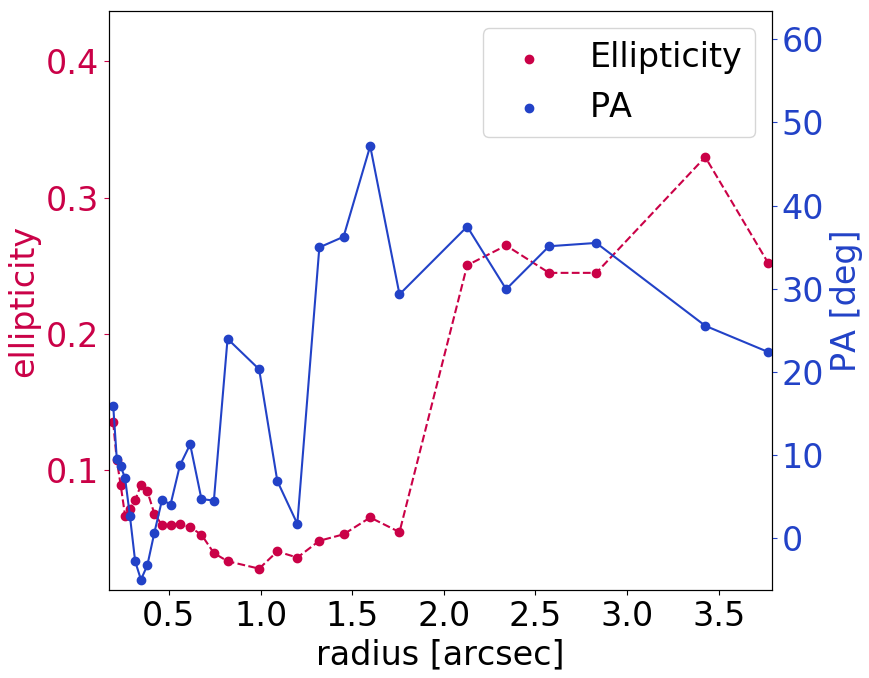}
\caption{SDSS J161259.83+421940.3. The ellipticity and the position angle plotted against the major axis of the isophote. Symbols and colours are explained in the plot.}
\label{fig:1612epa}
\end{figure}

Based on the PSF star the seeing was 0.8" during this observation. The four exposures we have are about the same quality, and the total exposure time is 3600s. Fitting a PSF and a S\'{e}rsic function 
leaves residuals (Fig.~\ref{fig:1612nonsym}) showing an excess of flux around the nuclear region -- possibly due to the circumnuclear star formation -- and a prominent brightening on the northwest 
side of the galaxy. This structure could be a spiral arm or a morphological disturbance due to a recent merger. We experimented with different combinations of functions to achieve as good a fit as 
possible: the best combination was a PSF and two S\'{e}rsic functions. Parameters of the best fit are presented in Table~\ref{tab:j1612}, the observed, model, and residual images of J161259.83+421940.3 
in Fig.~\ref{fig:1612}, and the radial surface brightness profiles in Fig.~\ref{fig:1612comps}. S\'{e}rsic 1 accounts for the bulge of the galaxy, which, based on the S\'{e}rsic index, is a pseudo-bulge.
Based on the fit parameters S\'{e}rsic 2 resembles a bar-like structure, but it is very faint as seen in its surface brightness profile. However, when plotting the ellipticity and the position angle 
against radius, a region with bar-like properties can be seen between approximately 1.7--3.5~arcsec. According to the best fit the bar component is quite extended with $r_e$=8.39~kpc, and there seems not 
to be a disk at all. However, the bar is very faint and the possible disk can be expected to be even fainter, in which case we probably would not be able to detect it with these observations. An 
alternative option is that the morphology is perturbed, as suggested by Fig.~\ref{fig:1612nonsym}, and instead of an undisturbed bar the feature seen in Fig.~\ref{fig:1612epa} traces a disturbed, 
elongated component caused by interaction. Both S\'{e}rsic functions are slightly off-centred from the PSF towards the bright feature, and together account for the excess flux. The contribution of 
the PSF to the total $J$-band flux is low, and the fit is almost as good without the PSF as it is with it. This indicates that the $J$-band emission is mostly dominated by the host galaxy, which 
seems plausible taking into account its enhanced star formation.

\begin{table*}
\caption[]{Best fit parameters of SDSS J161259.83+421940.3. $\chi^2_{\nu}$ = 1.11 $\substack{+0.01 \\ -0.00}$.}
\centering
\begin{tabular}{l l l l l l l}
\hline\hline
 funct        &  mag                              & $r_{e}$                          & $n$                              & axial                            & PA                   & notes   \\
              &                                   & (kpc)                            &                                  & ratio                            & (\textdegree)        &         \\ \hline
 PSF          & 20.09 $\substack{+0.11 \\ -0.11}$ &                                  &                                  &                                  &                                  &    \\
 S\'{e}rsic 1 & 16.90 $\substack{+0.10 \\ -0.11}$ & 1.19 $\substack{+0.00 \\ -0.00}$ & 0.64 $\substack{+0.00 \\ -0.00}$ & 0.89 $\substack{+0.00 \\ -0.00}$ & 63.2 $\substack{+0.00 \\ -0.00}$ & bulge \\
 S\'{e}rsic 2 & 18.20 $\substack{+0.20 \\ -0.21}$ & 8.39 $\substack{+0.30 \\ -0.22}$ & 0.32 $\substack{+0.08 \\ -0.06}$ & 0.65 $\substack{+0.01 \\ -0.00}$ & 23.5 $\substack{+0.9 \\ -0.6}$ & bar \\ \hline
\end{tabular}
\label{tab:j1612}
\end{table*}

\section{Discussion}
\label{sec:disc}

We successfully modelled the host galaxy morphologies of five NLS1 sources; the results are summarised in Table~\ref{tab:summary}. We were unable to model one source, J122749.14+321458.9, and the modelling of the remaining three galaxies, J123220.11+495721.8, J125635.89+500852.4 and J133345.47+414127.7, is probably compromised; these four sources suffered from poor seeing. Combined with some of the highest redshifts of the sample, the data obtained for these sources were not of good enough quality to properly model them. Their data are presented in appendix~\ref{sec:compromised-fits} 

All five reliably fitted sources are disk-like galaxies with black hole masses well below 10$^8 M_{\odot}$. Four of them have pseudo-bulges while the data of the remaining source were possibly not good enough for distinguishing all galaxy components from each other. The prevalence of pseudo-bulges supports the idea that NLS1 galaxies host black holes of low or intermediate masses \citep[e.g., ][]{2015jarvela1,2016cracco1}. Bars are also commonly seen, with four out of five sources showing a component that can be perceived as a bar.

The high fraction of sources that are interacting with another galaxy (J111934.01+533518.7 and J152205.41+393441.3) or showed signs of disturbed morphology during the fitting procedure (J161259.83+421940.3) is remarkable: three out of five of the properly modelled sources seem to be disturbed spiral galaxies. In addition one of the sources with worse quality data, J133345.47+414127.7, may show signs of interaction. However, since they in any case have the characteristics of late-type galaxies, for example, pseudo-bulges, it is improbable that they would previously have undergone major mergers. Two out of three have been detected at 37~GHz and we assume them to host powerful, most probably relativistic jets, and the third is a radio-loud NLS1 galaxy. This may indicate that morphological disturbances caused by galaxy -- galaxy interaction indeed are connected to higher levels of nuclear activity and subsequently the launching of relativistic jets in NLS1 galaxies. If the scenario proposed by \citet{2001mathur1} that NLS1 galaxies are sources rejuvenated by mergers is correct, the previous mergers must have been minor for the spiral-like morphology of the host galaxy to have been maintained.

Compared to radio-quiet and radio-silent NLS1 galaxies the sources in our sample seem to be quite similar: late-type galaxies with pseudo-bulges and with a high fraction of bars. The divisive factor is the frequency of interaction. Among radio-quiet and radio-silent NLS1 sources only 8\%-16\% show signs of interaction \citep{2007ohta1}, whereas among NLS1 galaxies with probable jets, when including the previously studied sources and excluding the four in our sample with the poor quality data, four (or six if assuming that FBQS J1644+2619 and PKS 2004-447 have gone through recent minor mergers) out of eight show disturbed morphology. The sample size is still small, but assuming that 16\% of NLS1 galaxies are interacting and we randomly choose eight sources, the probability that four of them are interacting is low (2.2\%). Thus it seems safe to say that among NLS1 galaxies the presence of a jet is most likely connected to interaction, but since there is also proof of isolated NLS1 galaxies with jets, for example, J151020.06+554722.0 in this study, it is not the only factor that influences nuclear activity.

Our results highlight the disparity between NLS1 galaxies with and without jets (assuming that 37~GHz emission is an indicator of relativistic jets), and further support their heterogeneity. If interaction, mostly minor mergers, indeed proves to be one of the dividing factors, this suggests that the observed heterogeneity of the NLS1 population might be due to their development over time. This also explains and supports the findings in \citet{2017jarvela1} where NLS1 galaxies with jets were found to favour denser large-scale environments than NLS1 galaxies without jets. The probability of interaction, which might lead to the launching of a jet, is naturally higher in regions with more galaxies. One out of the five reliably modelled sources in our sample resides in a supercluster, three in an intermediate-density region, and one in a void \citep[data from ][]{2017jarvela1}. The only source in supercluster environment, J111934.01+533518.7, shows signs of interaction. One of the three in the intermediate-density regions, J161259.83+421940.3, shows disturbed morphology. J091313.73+365817.2 seems to be an undisturbed spiral galaxy, but then again it has not been detected at 37~GHz and therefore probably does not host a jet. J151020.06+554722.0 is an intriguing source: it does not seem to be disturbed, and according to the SDSS Main Galaxy sample it is an isolated galaxy --- though it might have faint satellite galaxies not visible in the SDSS --- but it is one of the sources that has the highest detection rate at 37~GHz. This raises the question of what has triggered the high-frequency radio emission that we assume to originate in the jet. However, the seeing during the J151020.06+554722.0 observation was not optimal so it is possible that the signs of interaction have passed unnoticed. The source in a void, J152205.41+393441.3, is clearly a component in an ongoing merger, suggesting that it has a smaller companion galaxy it is now merging with. 

We find that q22 is not a good measure of the presence of a jet. In fact the two sources classified as exhibiting starburst activity possess jets based on 37~GHz observations and also show disturbed morphology. This parameter may correctly estimate the contribution of the various emission components around 1.4~GHz, but it does not seem to correlate with higher radio frequencies. It might well be that the emission is dominated by star formation processes at 1.4~GHz, especially if the galaxy exhibits (post-)merger starburst activity, but it does not mean that the source would not host a jet at the same time. Vice versa, starbursts are often triggered in a merger, as seems to be the case for jets as well. Since q22 identifies starburst sources as the most star-formation-dominated, these sources might also have a higher probability of exhibiting a jet. q22 uses VLA FIRST and WISE data which in the worst case could have been acquired decades apart, and it does not take into account the high variability of the AGN and jet emission at low radio frequencies. 

It is significant that none of the host galaxies of the NLS1 sources with assumed jets in our study is an elliptical galaxy (except the dubious case of J125635.89+500852.4). This result, combined with the other remarkable characteristics of NLS1 galaxies, for example, the unusually low black hole masses found also in this study, contradicts the usual AGN paradigm that only massive elliptical galaxies are able to launch and maintain relativistic jets.

\renewcommand{\arraystretch}{1.0}
\begin{table*}
\caption[]{Summary of the results.}
\centering
\begin{tabular}{l l l l l l}
\hline\hline
source              & morphology & components     & jets & large-scale  & notes   \\
                    &            &                &        & environment  &              \\ \hline
J091313.73+365817.2 & spiral     & PB, bar, disk  &        & intermediate & enhanced SF   \\ 
J111934.01+533518.7 & spiral     & PB             &        & supercluster & interacting, RL \\          
J122749.14+321458.9 & unclear    &                & yes    & intermediate & enhanced SF\\
J123220.11+495721.8 & unclear    &                & yes    & supercluster &   \\
J125635.89+500852.4 & unclear    &                & yes    & intermediate &   \\
J133345.47+414127.7 & unclear    &                & yes    & supercluster & starburst \\
J151020.06+554722.0 & spiral     & bar, disk      & yes    & intermediate &               \\
J152205.41+393441.3 & spiral     & PB, bar, disk  & yes    & void         & interacting             \\        
J161259.83+421940.3 & spiral     & PB, bar        & yes    & intermediate & disturbed, starburst \\ \hline
\end{tabular}
\tablefoot{Col. 1 gives the source name and Col. 2 the seeing during the observations. Col. 3 lists the probable morphology of the host galaxy and Col. 4 gives the components of the model; PB denotes pseudo-bulge. In Col. 5 a source is marked if it has been detected at 37~GHz. The large-scale environment parameters in Col. 6 are from \citet{2017jarvela1}. Col. 7 gives additional notes of the source; SF denotes star formation. }
\label{tab:summary}
\end{table*}

\section{Conclusions}
\label{sec:concl}

We observed nine NLS1 galaxies in near-infrared to determine the morphologies of their host galaxies. We were able to reliably model five sources of which at least three have jets based on 37~GHz observations. 
Our main conclusions are:

\begin{itemize}
    \item[$\bullet$] Five out of nine sources in our sample are late-type galaxies, most with pseudo-bulges and bars, and three of them host jets according to 37~GHz observations. Interaction and mergers are seen in three sources.
    \item[$\bullet$] The abundance of interacting systems may indicate that interaction is connected to nuclear activity and the launching of jets, and could explain the heterogeneity seen in the NLS1 population.
    \item[$\bullet$] As proof against the conventional view of relativistic jets being exclusively launched from massive elliptical galaxies accumulates, the current evolution and unification models need to be revised.
\end{itemize}

\clearpage

\begin{acknowledgements}

We are grateful to J. Harmanen for his help with the NOT spectral data reduction and analysis, and to FINCA Science School for obtaining the spectra. We thank R. De Propris for his help with GALFIT. Based on observations made with the Nordic Optical Telescope, operated by the Nordic Optical Telescope Scientific Association at the Observatorio del Roque de los Muchachos, La Palma, Spain, of the Instituto de Astrofisica de Canarias. This publication makes use of data obtained at the Mets\"{a}hovi Radio Observatory, operated by Aalto University, Finland. This work is based on observations made with the Copernico and Schmidt Telescopes of the INAF-Asiago Observatory. The National Radio Astronomy Observatory is a facility of the National Science Foundation operated under cooperative agreement by Associated Universities, Inc. This publication makes use of data products from the Wide-field Infrared Survey Explorer, which is a joint project of the University of California, Los Angeles, and the Jet Propulsion Laboratory/California Institute of Technology, funded by the National Aeronautics and Space Administration. This publication makes use of data products from the Two Micron All Sky Survey, which is a joint project of the University of Massachusetts and the Infrared Processing and Analysis Center/California Institute of Technology, funded by the National Aeronautics and Space Administration and the National Science Foundation. Funding for the Sloan Digital Sky Survey (SDSS) has been provided by the Alfred P. Sloan Foundation, the Participating Institutions, the National Aeronautics and Space Administration, the National Science Foundation, the U.S. Department of Energy, the Japanese Monbukagakusho, and the Max Planck Society. The SDSS Web site is http://www.sdss.org/. The SDSS is managed by the Astrophysical Research Consortium (ARC) for the Participating Institutions. The Participating Institutions are The University of Chicago, Fermilab, the Institute for Advanced Study, the Japan Participation Group, The Johns Hopkins University, the Korean Scientist Group, Los Alamos National Laboratory, the Max-Planck-Institute for Astronomy (MPIA), the Max-Planck-Institute for Astrophysics (MPA), New Mexico State University, University of Pittsburgh, University of Portsmouth, Princeton University, the United States Naval Observatory, and the University of Washington. 

\end{acknowledgements}

\bibliographystyle{aa}
\bibliography{artikkeli.bib}

\begin{appendix}

\section{Compromised fits}
\label{sec:compromised-fits}

The following sources did not have good enough data to properly model them. These sources suffered a combination of some of the highest redshifts among our sample and worst weather conditions during the observations. Only the radial surface brightness profile of the galaxy are presented. The parameters of the fits are not stable but vary considerably depending on the sky value, and are therefore not reliable. The host galaxy morphologies of these sources remain unclear.

\subsection{SDSS J122749.14+321458.9}

J122749.14+321458.9 is classified as a radio-loud flat-spectrum NLS1 galaxy. It has been detected at 37~GHz but at a rather low flux density level of around 0.2~Jy. q22 of 1.19 and W3-W4 of 2.62 indicate strong star formation; 54 $M_{\odot}$ yr$^{-1}$ according to \citet{2015caccianiga1}. In the JVLA radio map J122749.14+321458.9 has a compact, unresolved core and possibly some diffuse radio emission on one side of the source \citep{2018berton1}.

\begin{figure*}[ht!]
\centering
\adjustbox{valign=t}{\begin{minipage}{0.36\textwidth}
\centering
\includegraphics[width=1\textwidth]{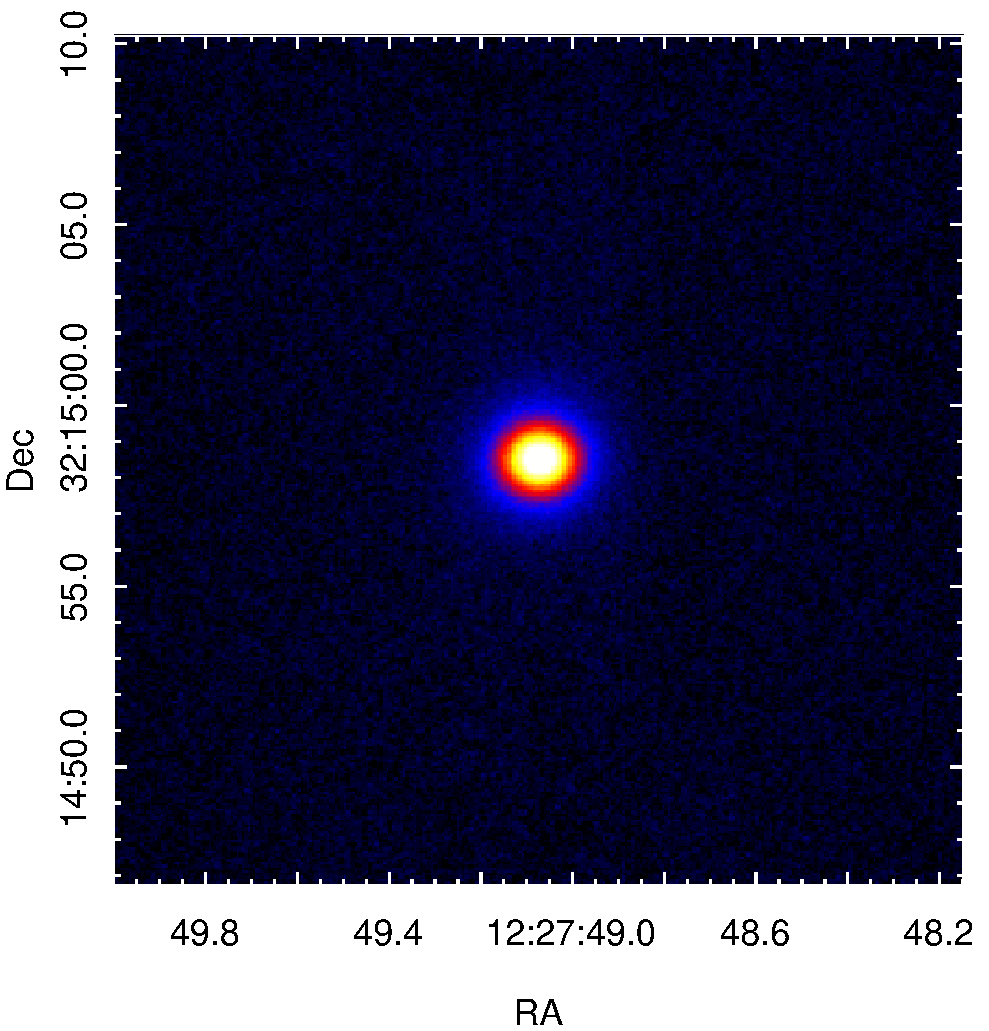}
\end{minipage}}
\adjustbox{valign=t}{\begin{minipage}{0.31\textwidth}
\centering
\includegraphics[width=1\textwidth]{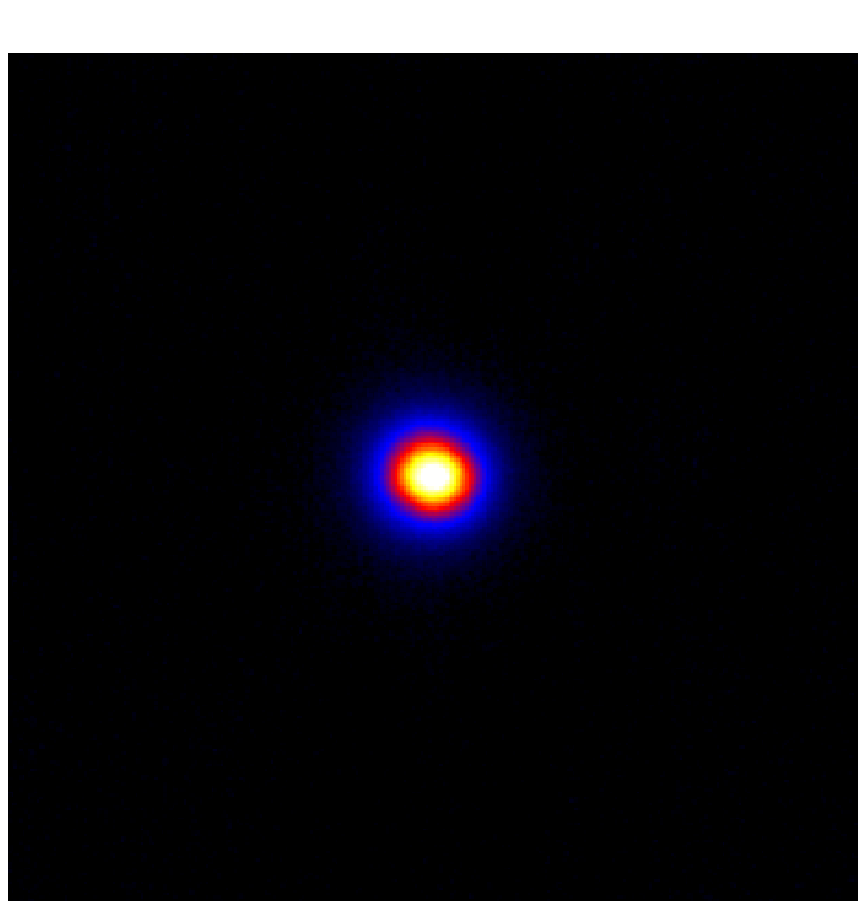}
\end{minipage}}
\adjustbox{valign=t}{\begin{minipage}{0.31\textwidth}
\centering
\includegraphics[width=1\textwidth]{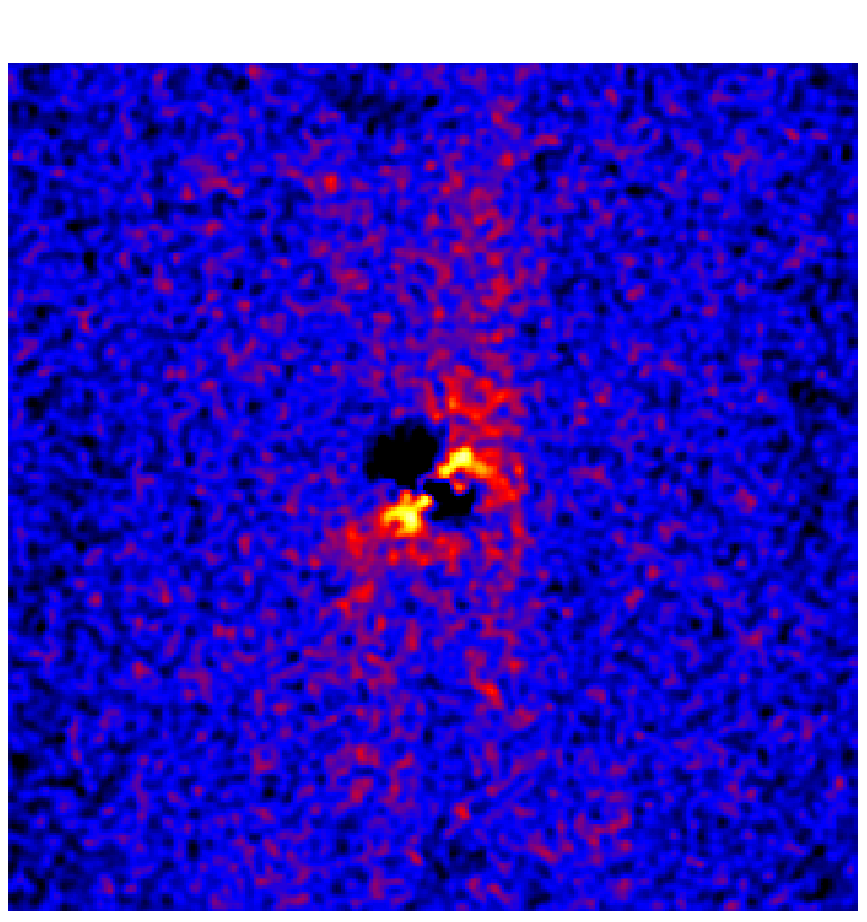}
\end{minipage}}
\hfill
    \caption{SDSS J122749.14+321458.9. The field of view is 23.4" / 54.5~kpc in all images. \emph{Left panel:} observed image, \emph{middle panel:} model image, and \emph{right panel:} residual image, smoothed over 3px.} \label{fig:1227}
\end{figure*}

\begin{figure}
\centering
\includegraphics[width=0.5\textwidth]{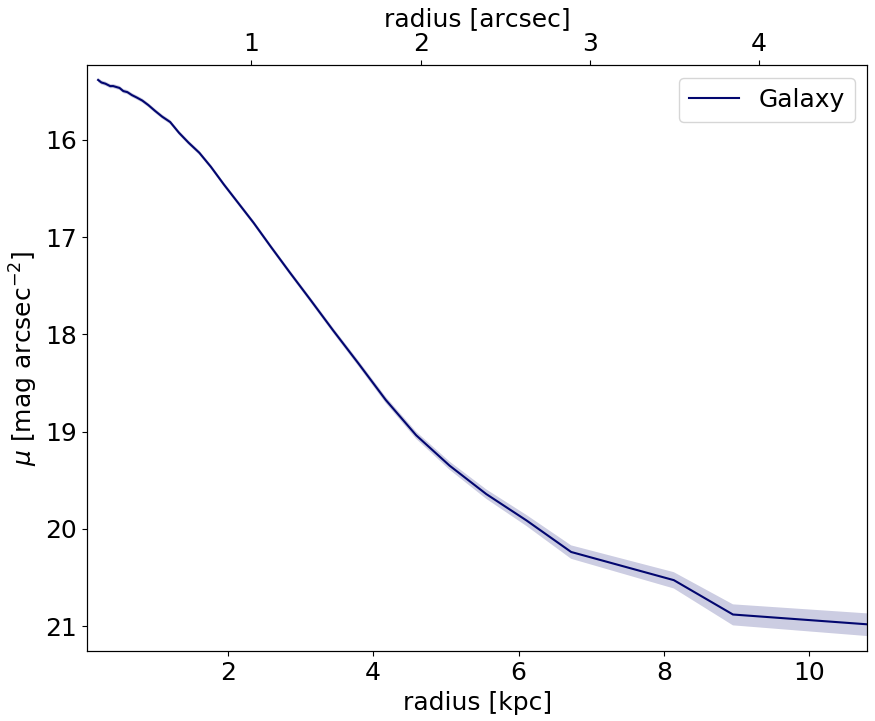}
\caption{SDSS J122749.14+321458.9. The observed radial surface brightness profile. Symbols and colours are explained in the plot. The shaded area around the profile describes the associated errors.}
\label{fig:1227comps}
\end{figure}

We have two 900s exposures of this source with seeing of 1.2". The fit with only a PSF left prominent axisymmetric residuals, probably due to compromised PSF modelling and poor quality data. We attempted to model the source with the PSF and a variety of other functions and their combinations, but in all cases the functions converged to parameters that are not physical; $n$ around 15--20 and the axial ratio betwen 0.01--0.05. It seems that due to the poor seeing, combined with the short exposure time, J122749.14+321458.9 cannot be properly modelled. The fit is shown only with the PSF. The parameters are given in Table~\ref{tab:j1227}, the observed, model, and residual images in Fig.~\ref{fig:1227}, and the radial surface brightness profile of the galaxy in Fig.~\ref{fig:1227comps}. 

\renewcommand{\arraystretch}{1.5}
\begin{table}
\caption[]{Best fit parameters of SDSS J122749.14+321458.9. $\chi^2_{\nu}$ = 1.22 $\substack{+0.02 \\ -0.01}$.}
\centering
\begin{tabular}{l l l l l l l l}
\hline\hline
function   &  Mag  & $r_{e}$ & $n$  & axial & PA    & notes   \\
           &       & (kpc)   &      & ratio & (\textdegree)  &  \\ \hline
PSF        & 15.11 $\substack{+0.10 \\ -0.10}$ &         &      &       &       &   \\  \hline
\end{tabular}
\label{tab:j1227}
\end{table}

\subsection{SDSS J123220.11+495721.8}

J123220.11+495721.8 is a source previously classified radio-silent. It has, however, been detected at 37~GHz multiple times with a maximum flux density of 0.56~Jy. Its star formation indicators are not particularly high, q22 = 0.94 (assuming $S_{\mathrm{1.4~GHz}}$ = 1~mJy) and W3-W4 = 2.19, however, these are lower limits and in reality they can be higher.

\begin{figure*}[ht!]
\centering
\adjustbox{valign=t}{\begin{minipage}{0.34\textwidth}
\centering
\includegraphics[width=1\textwidth]{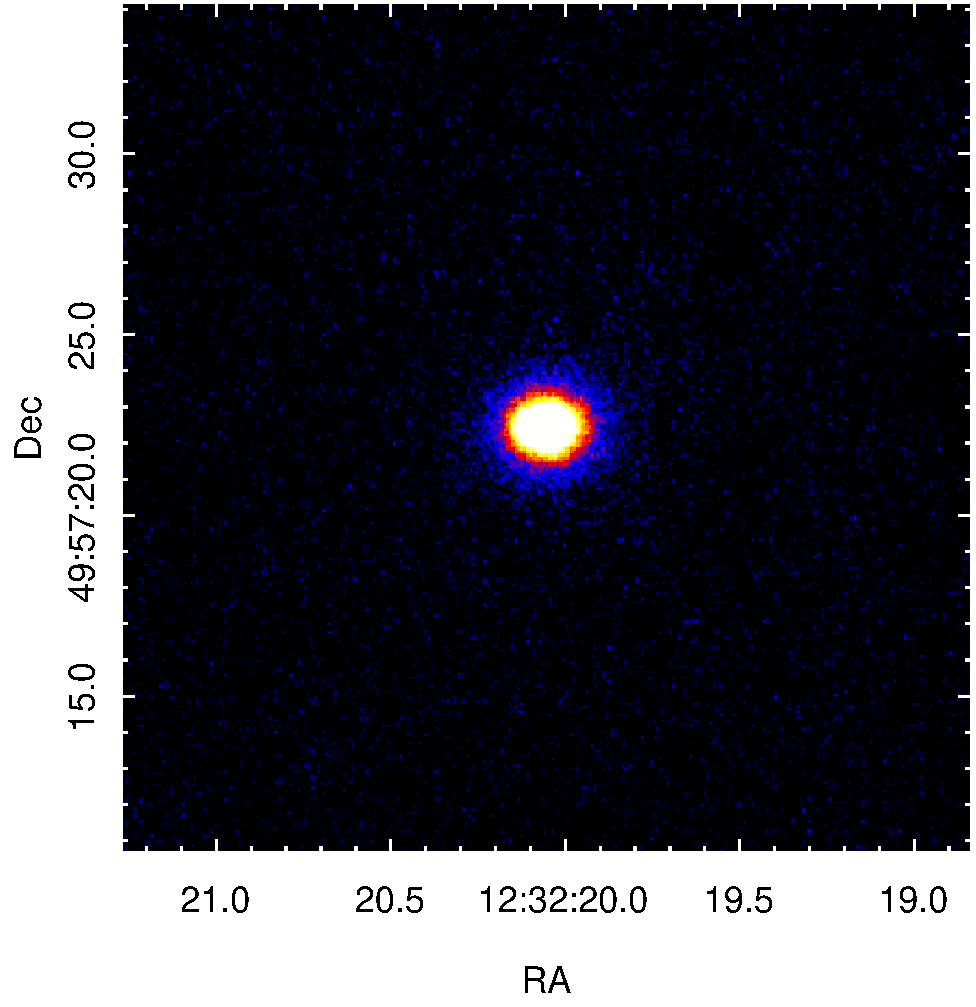}
\end{minipage}}
\adjustbox{valign=t}{\begin{minipage}{0.32\textwidth}
\centering
\includegraphics[width=0.95\textwidth]{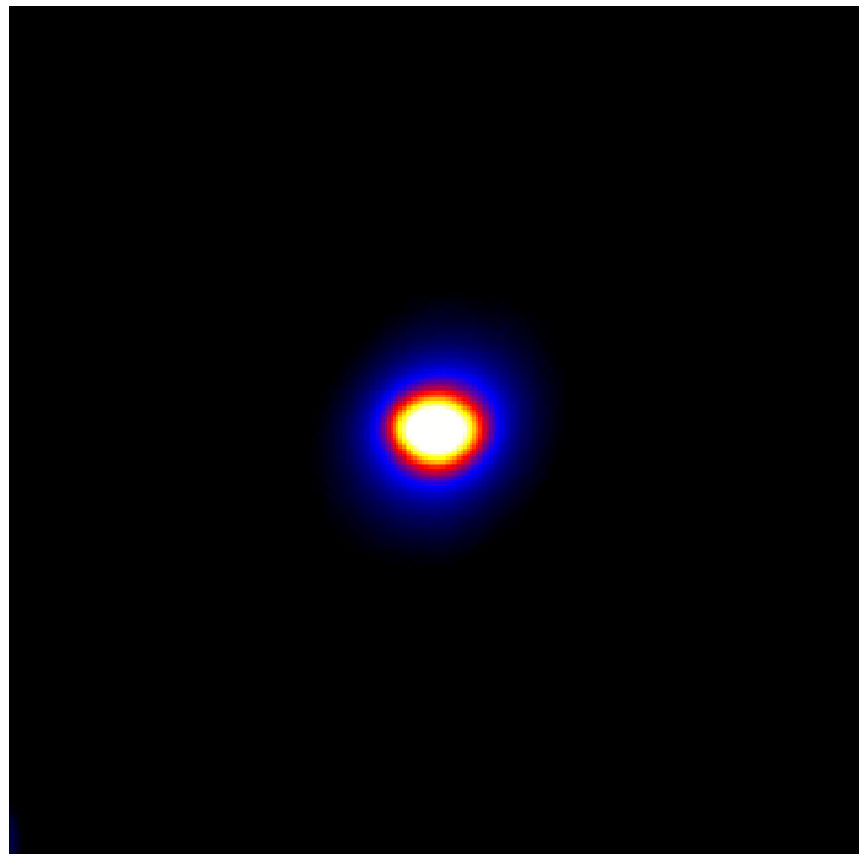}
\end{minipage}}
\adjustbox{valign=t}{\begin{minipage}{0.32\textwidth}
\centering
\includegraphics[width=0.95\textwidth]{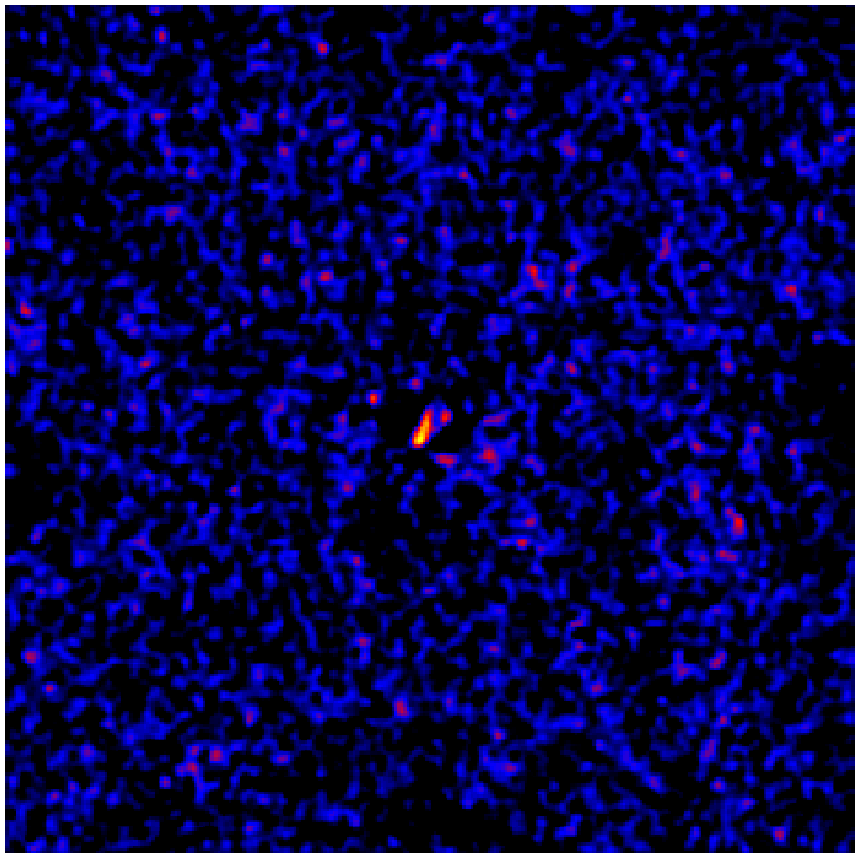}
\end{minipage}}
\hfill
    \caption{SDSS J123220.11+495721.8. The field of view is 23.4" / 91.3~kpc in all images. \emph{Left panel:} observed image, \emph{middle panel:} model image, and \emph{right panel:} residual image, smoothed over 3px.} \label{fig:1232}
\end{figure*}

\begin{figure}
\centering
\includegraphics[width=0.5\textwidth]{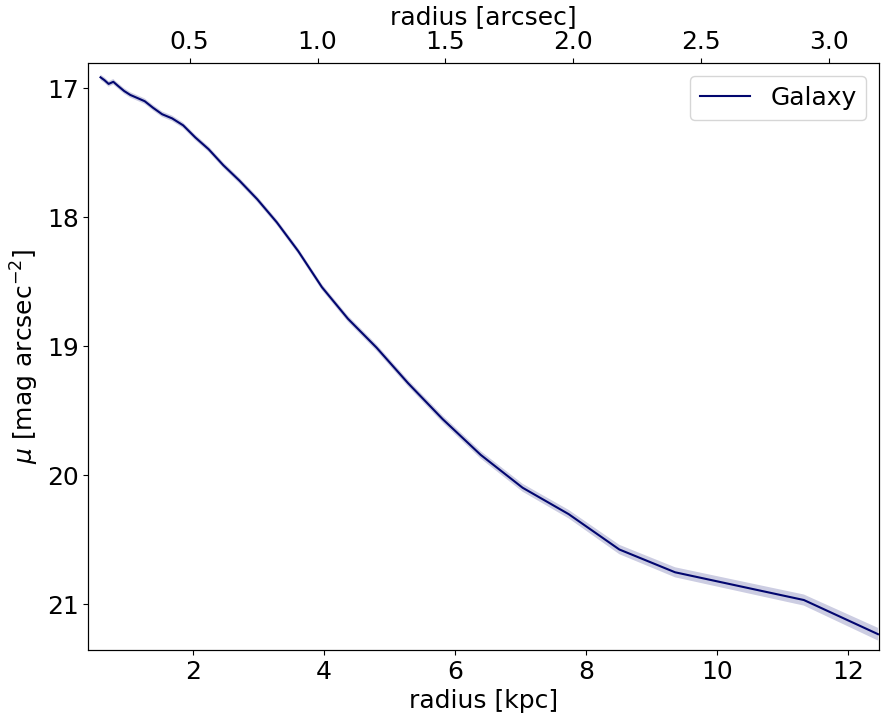}
\caption{SDSS J123220.11+495721.8. The observed radial surface brightness profile. Symbols and colours are explained in the plot. The shaded area around the profile describes the associated errors.}
\label{fig:1232comps}
\end{figure}

The exposure time for this source was 3600s and the seeing $\sim$1". The source is fitted with a PSF and a S\'{e}rsic. The parameters of the fit are listed in Table~\ref{tab:j1232}, Fig.~\ref{fig:1232} shows the observed, model, and residual images, and Fig.~\ref{fig:1232comps} the radial surface brightness profile of the galaxy. The S\'{e}rsic index is suspiciously high and depends on the sky value, making the fit physically unreliable. In fact, the fit seems to be equally good with \emph{any} S\'{e}rsic index between 1--20, adding to the inaccuracy of the result. 

\begin{table*}
\caption[]{Best fit parameters of SDSS J123220.11+495721.8. $\chi^2_{\nu}$ = 1.17 $\substack{+0.00 \\ -0.00}$.}
\centering
\begin{tabular}{l l l l l l l l}
\hline\hline
function   &  mag                              & $r_{e}$                          & $n$                               & axial                            & PA    & notes   \\
           &                                   & (kpc)                            &                                   & ratio                            & (\textdegree)  &  \\ \hline
PSF        & 18.21 $\substack{+0.50 \\ -0.10}$ &                                  &                                   &                                  &       &   \\
S\'{e}rsic & 16.54 $\substack{+0.10 \\ -0.19}$ & 3.90 $\substack{+0.00 \\ -1.08}$ & 18.37 $\substack{+1.58 \\ -0.00}$ & 0.74 $\substack{+0.00 \\ -0.00}$ & 55.1 $\substack{+0.0 \\ -2.3}$ & \\
S\'{e}rsic & 19.16 $\substack{+0.15 \\ -0.16}$ & 2.81 $\substack{+0.10 \\ -0.06}$ & 0.61 $\substack{+0.19 \\ -0.11}$  & 0.87 $\substack{+0.01 \\ -0.02}$ & -78.7 $\substack{+1.5 \\ -0.2}$ & nearby \\  \hline
\end{tabular}
\label{tab:j1232}
\end{table*}

\subsection{SDSS J125635.89+500852.4}

J125635.89+500852.4 is a source previously classified as radio-silent, however, it has been detected at 37~GHz with a flux density of $S_{\mathrm{37~GHz, max}}$ = 0.61~Jy. Assuming a 1.4~GHz flux density of 1~mJy its q22 = 0.61 and W3-W4 = 2.43, therefore the jet does not seem to be dominant at low radio frequencies.

\begin{figure*}[ht!]
\centering
\adjustbox{valign=t}{\begin{minipage}{0.34\textwidth}
\centering
\includegraphics[width=1\textwidth]{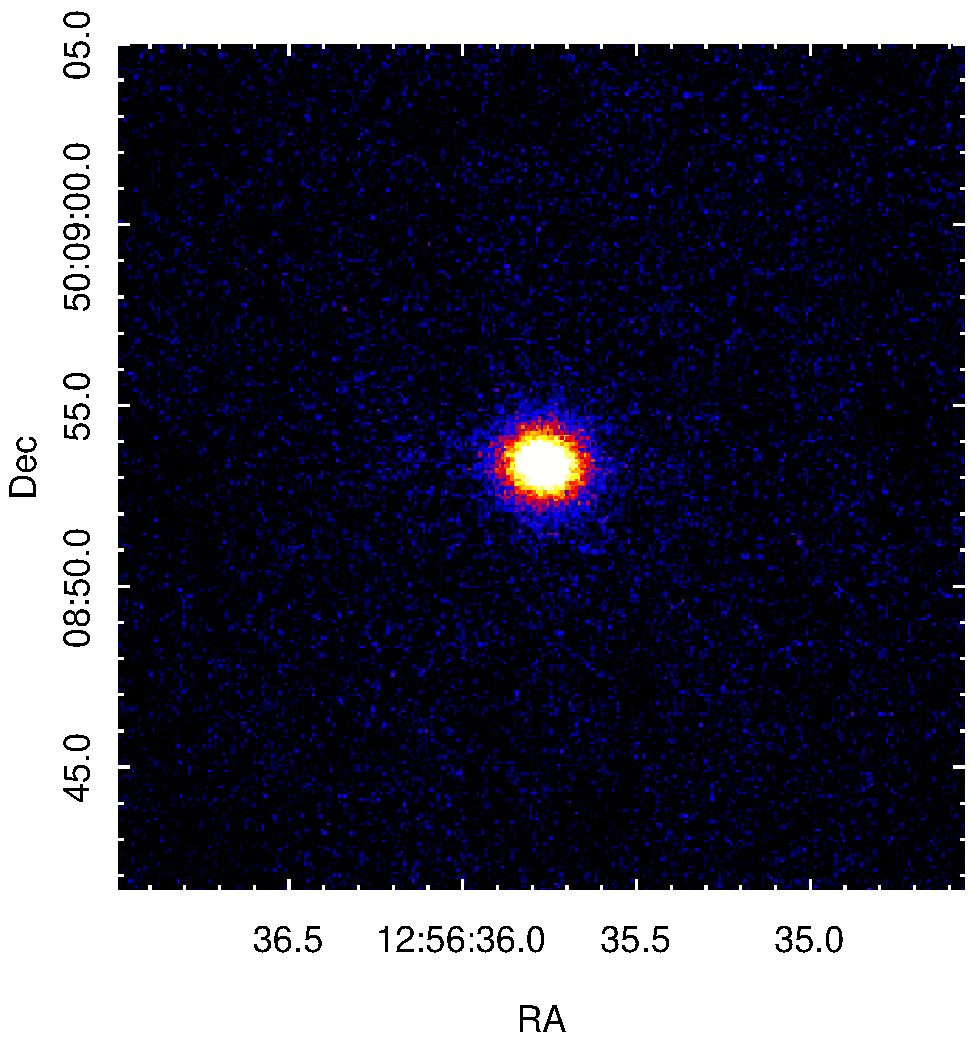}
\end{minipage}}
\adjustbox{valign=t}{\begin{minipage}{0.32\textwidth}
\centering
\includegraphics[width=0.95\textwidth]{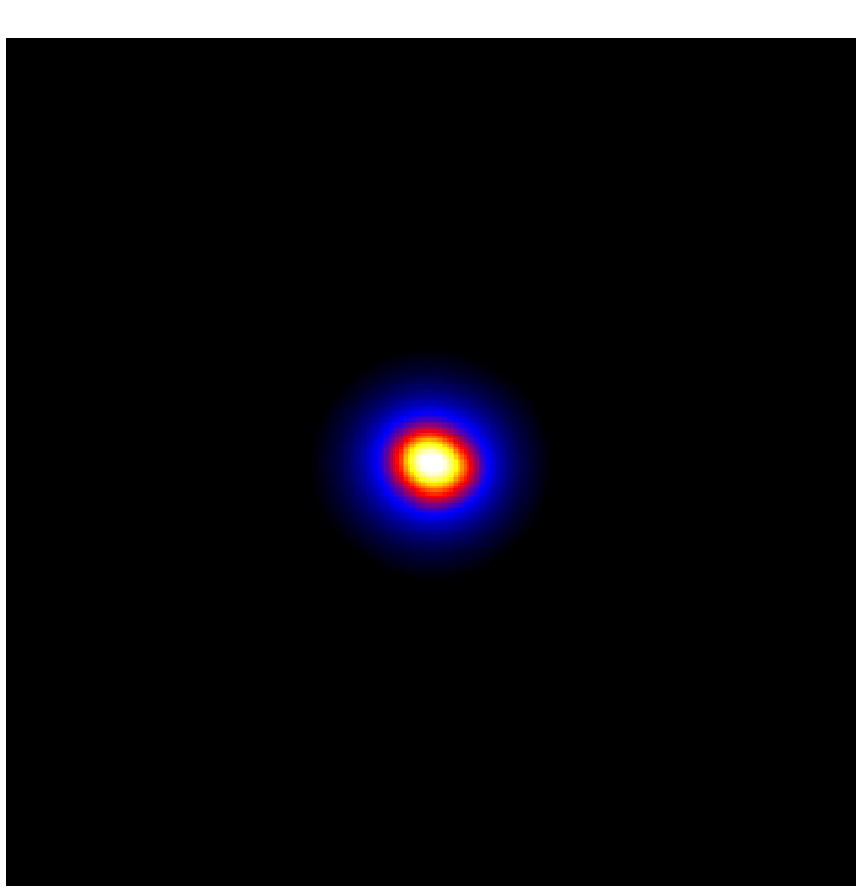}
\end{minipage}}
\adjustbox{valign=t}{\begin{minipage}{0.32\textwidth}
\centering
\includegraphics[width=0.95\textwidth]{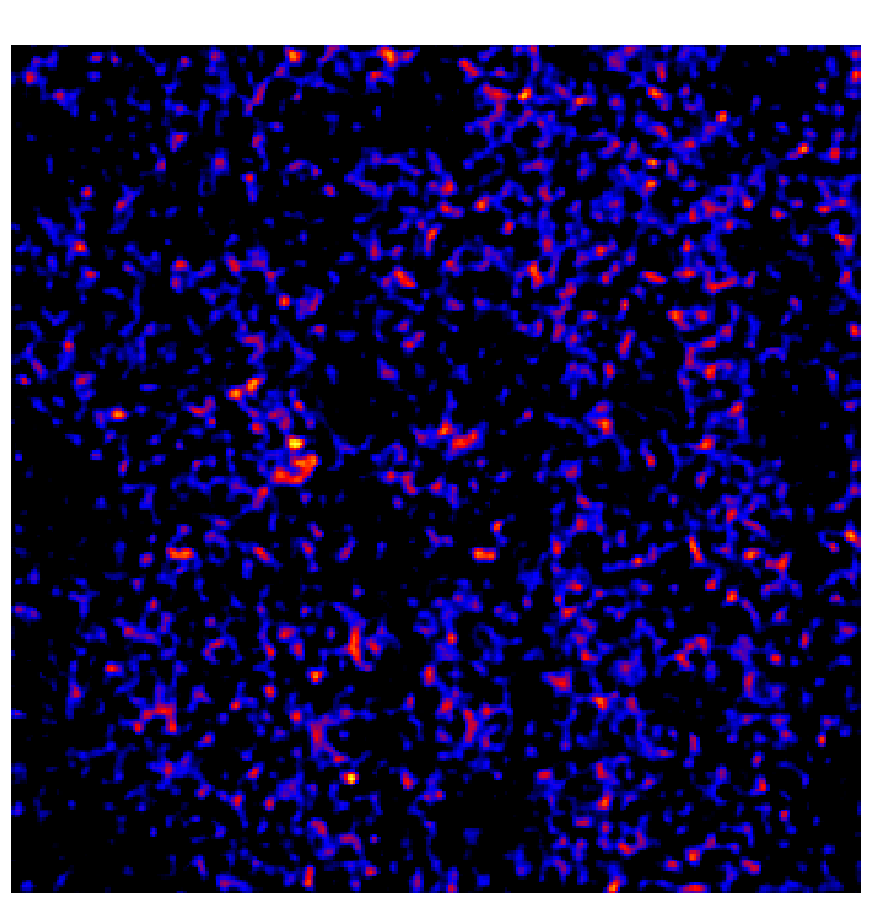}
\end{minipage}}
\hfill
    \caption{SDSS J125635.89+500852.4. The field of view is 23.4" / 87.0~kpc in all images. \emph{Left panel:} observed image, \emph{middle panel:} model image, and \emph{right panel:} residual image, smoothed over 3px.} \label{fig:1256}
\end{figure*}

\begin{figure}
\centering
\includegraphics[width=0.5\textwidth]{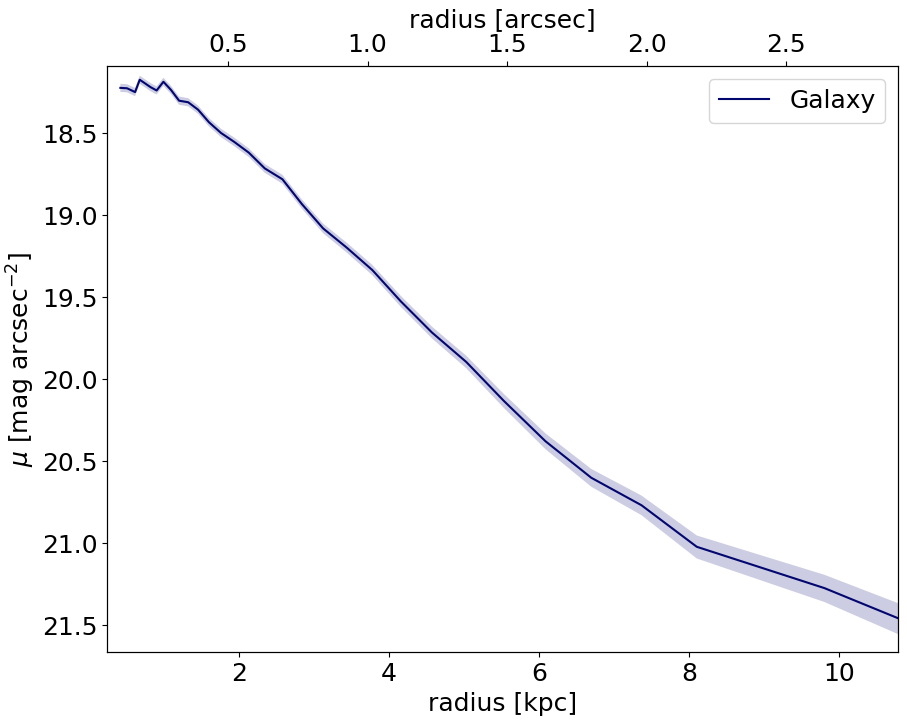}
\caption{SDSS J125635.89+500852.4. The observed radial surface brightness profile. Symbols and colours are explained in the plot. The shaded area around the profile describes the associated errors.}
\label{fig:1256comps}
\end{figure}

Seeing during the PSF star observation was 0.9", but got worse ( $\sim$1.4") during the source observation. Since the source lies at a considerable redshift, the bad seeing is likely to affect the results of the modelling. We have four 900s exposures of this source. We modelled the sources with a PSF and a S\'{e}rsic function, however, the fit parameters are not stable. Table~\ref{tab:j1256} lists the fit parameters, the observed, model, and residual images are shown in Fig.~\ref{fig:1256}, and the radial surface brightness profile of the galaxy in Fig.~\ref{fig:1256comps}. The S\'{e}rsic index of the model is 5.4, suggesting that J125635.89+500852.4 is an elliptical galaxy. However, the S\'{e}rsic index as well as other model parameters considerably change with the 1$\sigma$ values of the sky, casting doubt on the credibility of the model. Interestingly, the nucleus of J125635.89+500852.4 is faint compared to the galaxy and indeed, the fit is equally good without the PSF. 

\begin{table*}
\caption[]{Best fit parameters of SDSS J125635.89+500852.4. $\chi^2_{\nu}$ = 1.18 $\substack{+0.01 \\ -0.00}$.}
\centering
\begin{tabular}{l l l l l l l l}
\hline\hline
function   &  Mag                              & $r_{e}$                           & $n$                              & axial  & PA & notes  \\
           &                                   & (kpc)                             &                                  & ratio  & (\textdegree) &    \\ \hline
PSF        & 20.70 $\substack{+0.15 \\ -0.47}$ &                                   &                                  &        &       &  \\
S\'{e}rsic & 17.20 $\substack{+0.45 \\ -0.17}$ & 5.21 $\substack{+49.67 \\ -2.53}$ & 6.32 $\substack{+8.74 \\ -3.40}$ & 0.93 $\substack{+0.05 \\ -0.06}$ & -6.43 $\substack{+0.0 \\ -1.2}$ & \\  \hline
\end{tabular}
\label{tab:j1256}
\end{table*}

\subsection{SDSS J133345.47+414127.7} 

J133345.47+414127.7 is a borderline source between radio-quiet and radio-loud. It was previously classified as a flat-spectrum source, but based on JVLA observations re-classified as a steep-spectrum source with compact radio morphology \citep{2018berton1}. However, detections at 37~GHz indicate that it has a strongly variable jet. Star formation tracers of J133345.47+414127.7 are high, q22 = 1.3 and W3-W4 = 2.9. \citet{2015caccianiga1} estimate its SFR to be as high as 148 M$_{\odot}$ yr$^{-1}$.

\begin{figure}
\centering
\includegraphics[width=0.5\textwidth]{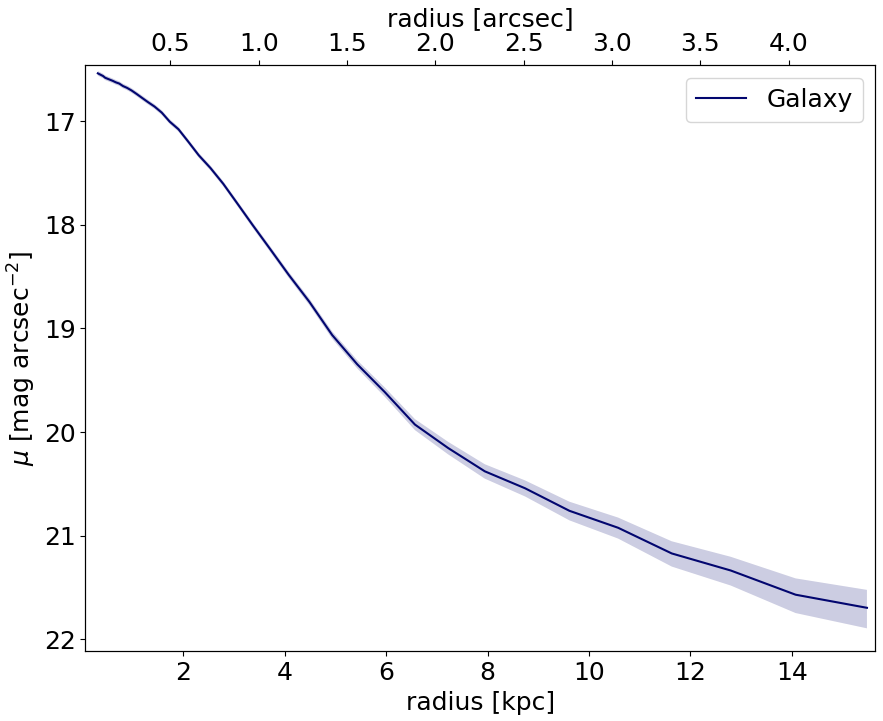}
\caption{SDSS J133345.47+414127.7. The observed radial surface brightness profile. Symbols and colours are explained in the plot. The shaded area around the profile describes the associated errors.}
\label{fig:1333comps}
\end{figure}

\begin{figure*}[ht!]
\centering
\adjustbox{valign=t}{\begin{minipage}{0.35\textwidth}
\centering
\includegraphics[width=1\textwidth]{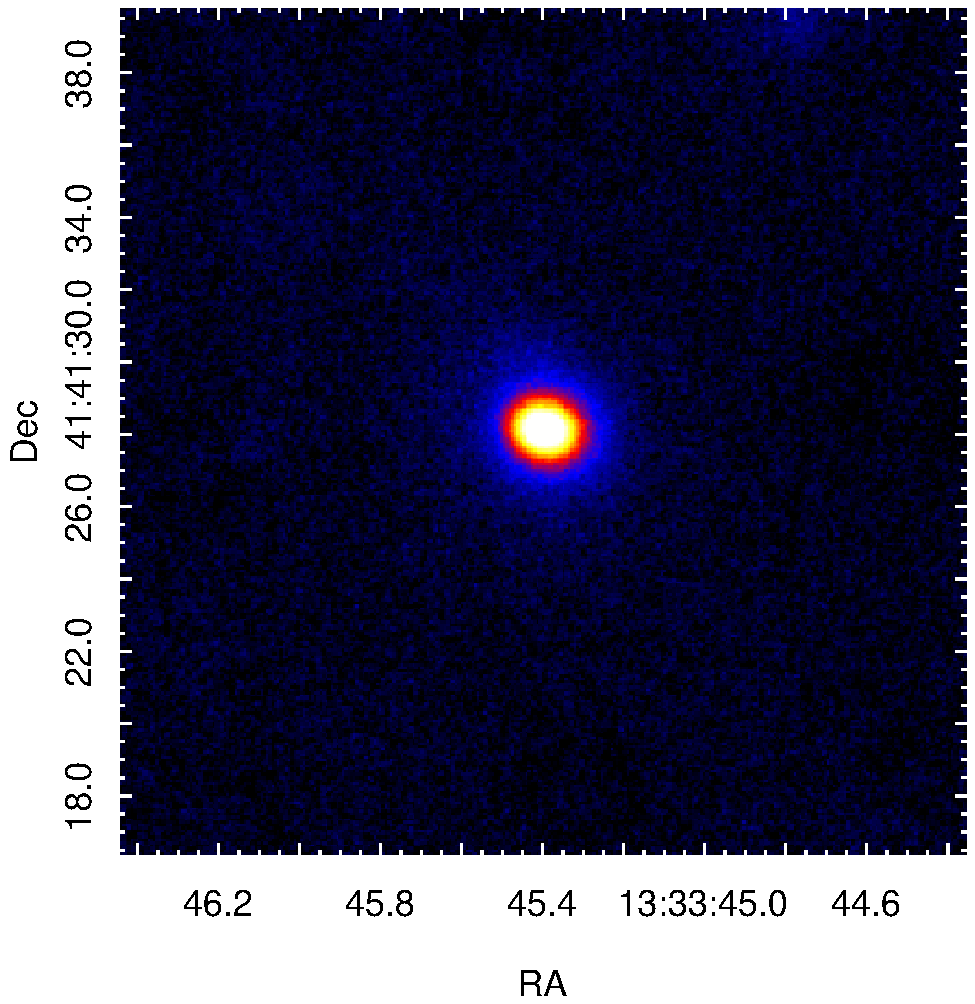}
\end{minipage}}
\adjustbox{valign=t}{\begin{minipage}{0.32\textwidth}
\centering
\includegraphics[width=0.95\textwidth]{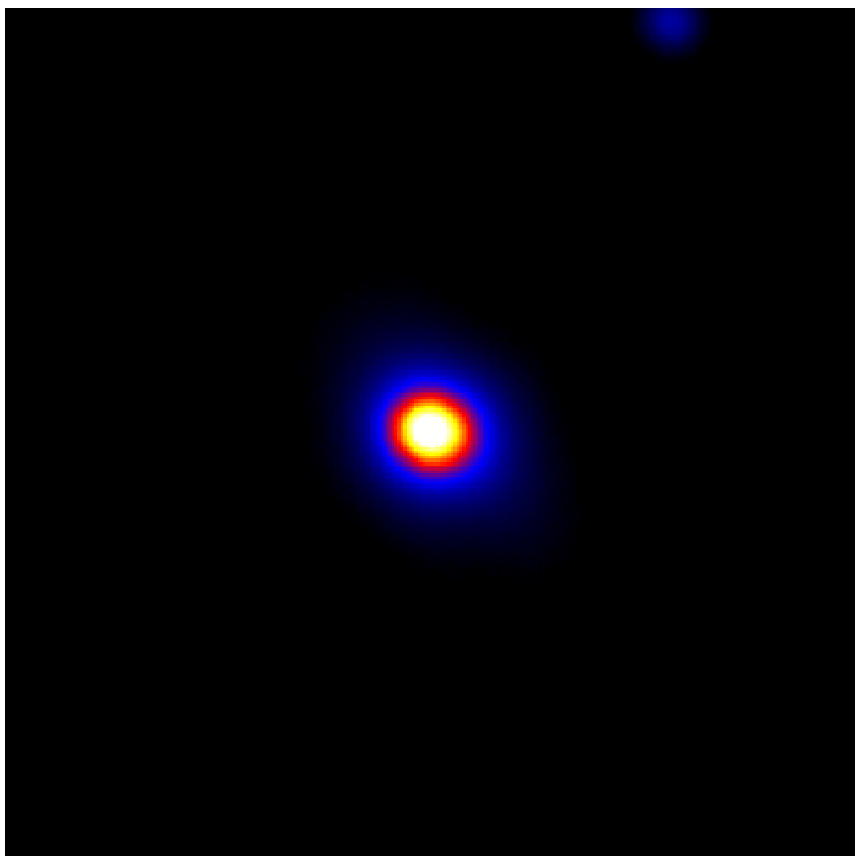}
\end{minipage}}
\adjustbox{valign=t}{\begin{minipage}{0.32\textwidth}
\centering
\includegraphics[width=0.95\textwidth]{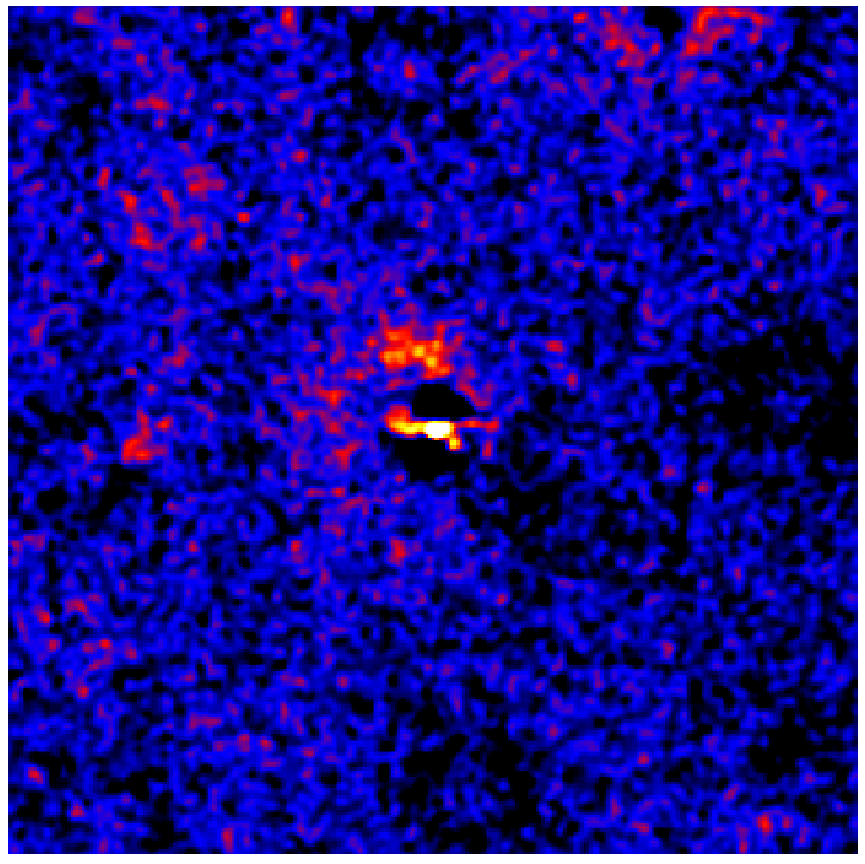}
\end{minipage}}
\hfill
    \caption{SDSS J133345.47+414127.7. The field of view is 23.4" / 81.5~kpc in all images. \emph{Left panel:} observed image, \emph{middle panel:} model image, and \emph{right panel:} residual image, smoothed over 3px.} \label{fig:1333}
\end{figure*}

Based on the PSF star the seeing was not optimal: $\sim$1.1". We have four 900s exposures of approximately the same quality. To increase the fitting area we also fitted two nearby sources, 
which were sufficiently modelled with PSFs. After several attempts the best fit of J133345.47+414127.7 was achieved with two PSFs and a S\'{e}rsic function. Trying to fit the galaxy with only one PSF, 
and a single or multiple S\'{e}rsic functions consistently yielded results with one of the S\'{e}rsic functions having an index of $\sim$20 that is not realistic, however, it indicates the presence of 
an additional compact component. Adding another PSF solves this problem. The separation of the centres of PSFs is 2.25~kpc. This is less than what the seeing allows to resolve so the fit should be 
treated with caution. For this reason the best fit parameters of the fit with a single and with two PSFs are shown in Tables~\ref{tab:j1333} and \ref{tab:j1333-2}, respectively. The observed, 
model, and residual images of the fit with one PSF and a S\'{e}rsic function are presented in Fig.~\ref{fig:1333}, and the radial surface brightness profile of the galaxy in Fig.~\ref{fig:1333comps}. 
Looking at the values and errors of the fits it is clear that they are not stable and most probably not reliable. However, in the residual image in Fig.~\ref{fig:1333}, even if the fit of the 
nuclear region is bad, it is evident that there is nonaxisymmetric flux excess on the northern side of the galaxy that could be a tidal feature.

\begin{table*}
\caption[]{Best fit parameters of SDSS J133345.47+414127.7, one PSF. $\chi^2_{\nu}$ = 1.14 $\substack{+0.02 \\ -0.00}$}
\centering
\begin{tabular}{l l l l l l l l}
\hline\hline
function   & Mag                               & $r_{e}$                             & $n$                                & axial                            & PA            & notes   \\
           &                                   & (kpc)                               &                                    & ratio                            & (\textdegree) &  \\ \hline
PSF        & 16.61 $\substack{+0.10 \\ -0.18}$ &                                     &                                    &                                  &               &         \\
S\'{e}rsic & 16.58 $\substack{+1.35 \\ -3.00}$ & 54.81 $\substack{^a \\ -49.73}$    & 15.16 $\substack{+0.00 \\ -12.47}$ & 0.42 $\substack{+0.10 \\ -0.00}$ & -47.9 $\substack{+4.6 \\ -1.6}$ & \\
\end{tabular}
\tablefoot{(a) The maximum value was not physically meaningful.}
\label{tab:j1333}
\end{table*}

\begin{table*}
\caption[]{Best fit parameters of SDSS J133345.47+414127.7, two PSFs. $\chi^2_{\nu}$ = 1.13 $\substack{+0.01 \\ -0.00}$}
\centering
\begin{tabular}{l l l l l l l l}
\hline\hline
function   & Mag                               & $r_{e}$                              & $n$                              & axial                            & PA            & notes   \\
           &                                   & (kpc)                                &                                  & ratio                            & (\textdegree) &  \\ \hline
PSF        & 16.56 $\substack{+0.20 \\ -0.14}$ &                                      &                                  &         &               &         \\
PSF        & 18.93 $\substack{+0.24 \\ -0.72}$ &                                      &                                  &                                  &               &         \\
S\'{e}rsic & 17.77 $\substack{+0.68 \\ -4.29}$ & 13.38 $\substack{^a \\ -4.58}$      & 1.01 $\substack{+5.38 \\ -0.61}$ & 0.52 $\substack{+0.15 \\ -0.03}$ & -50.7 $\substack{+1.5 \\ -8.0}$ & disk  \\
\end{tabular}
\tablefoot{(a) The maximum value was not physically meaningful.}
\label{tab:j1333-2}
\end{table*}

\end{appendix}

\end{document}